\documentclass[%
 aip,
groupedaddress,
 amsmath,amssymb,
preprint,
]{revtex4-1}

\usepackage[colorlinks=true,linkcolor=blue]{hyperref}
\usepackage{graphicx}% Include figure files
\usepackage{dcolumn}% Align table columns on decimal point
\usepackage{bm}% bold math
\usepackage[utf8]{inputenc}
\usepackage[T1]{fontenc}
\usepackage{mathptmx}
\usepackage{etoolbox}
\usepackage{multirow}
\usepackage{rotating}
\usepackage{braket}
\usepackage{bm}
\usepackage{float}
\usepackage[version=4]{mhchem}
\usepackage{booktabs}
\usepackage{calc}
\usepackage{siunitx}

\usepackage{pgfplots}
\makeatletter
\def\@email#1#2{%
 \endgroup
 \patchcmd{\titleblock@produce}
  {\frontmatter@RRAPformat}
  {\frontmatter@RRAPformat{\produce@RRAP{*#1\href{mailto:#2}{#2}}}\frontmatter@RRAPformat}
  {}{}
}%
\makeatother
\begin{document}

\preprint{AIP/123-QED}

\title[]{ Beyond the Dailey-Townes model: chemical information from the electric field gradient.}
% Force line breaks with \\
\author{G. Fabbro}
%\email{gfabbro@irsamc.upt-tlse.fr}
\author{J. Pototschnig}
%\email{jpoto@irsamc.ups-tlse.fr}
\author{T. Saue}\email{trond.saue@irsamc.ups-tlse.fr}
\homepage{https://dirac.ups-tlse.fr/saue}
\affiliation{Laboratoire de Chimie et Physique Quantique,\\UMR 5626 CNRS - Université Toulouse III-Paul Sabatier,\\ 118 Route de Narbonne, F-31062 Toulouse, France}

\date{\today}% It is always \today, today,

\begin{abstract}

In this work, we reexamine the Dailey--Townes model by systematically investigating the electric field gradient (EFG) in   various chlorine compounds, dihalogens, and the uranyl ion           (UO$_2^{2+}$). Through the use of relativistic molecular calculations and projection analysis, we decompose the EFG expectaton value in terms of atomic reference orbitals. We show how the Dailey--Townes model can be seen as an approximation to our projection analysis. Moreover, we observe  for the chlorine compounds that, in general, the Dailey-Townes model deviates from the total EFG value. We show that the main reason for this is that the Dailey--Townes model does not account for contributions from the mixing of valence $p$-orbitals with subvalence ones. We also find a non-negligible contribution from core polarization. This can be interpreted as Sternheimer shielding, as discussed in an appendix. The predictions of the Dailey--Townes model are improved by replacing net populations by gross ones, but we have not found any theoretical justification for this. Subsequently, for the molecular systems X–Cl (where X = I, At, and Ts), we find that with the inclusion of spin-orbit interaction, the (electronic) EFG operator is no longer diagonal within an atomic shell, which is incompatible with the Dailey--Townes model. Finally, we examine the EFG at the uranium position in UO$_2^{2+}$, where we find that about half the EFG comes from core polarization. The other half comes from the combination of the \ce{U\equiv O} bonds and the U$(6p)$ orbitals, the latter mostly non-bonding, in particular with spin-orbit interaction included. The analysis was carried out with molecular orbitals localized according to the Pipek–Mezey criterion. Surprisingly, we observed that core orbitals are also rotated during this localization procedure, even though they are fully localized. We show in an appendix that, using this localization criterion, it is actually allowed.

\end{abstract}

\maketitle

\section{\label{sec:level1}Introduction}
In 1935 Schüler and Schmidt\cite{Schuler_ZP1935a} reported clear deviations from the Landé interval rule, associated with the nuclear magnetic dipole moment, in the hyperfine splittings of the atomic spectra of the two naturally occurring isotopes of europium, \ce{^{151}Eu} and \ce{^{153}Eu}, both with nuclear spin I=5/2. They argued that these perturbations were due to a nuclear property beyond the little that was known at the time (mass, charge, mechanical and magnetic moment as well as volume (isotope shift)), namely a non-spherical nuclear charge distribution.\cite{Brix_ZNA1986} They assumed an ellipsoid shape, and noted in a subsequent publication, citing Delbr{\"u}ck,  that this corresponds to a quadrupole.\cite{Schueler_ZfP1935b,Smith_CSE1986}. Soon thereafter the underlying theory was worked out by Casimir.\cite{Casimir_1935,Casimir_book1936}

Nuclear electric quadrupole moments (NQMs) have played an important role in our understanding  of nuclear structure.  Upon inspection of quadrupole moments known at the time (1949), Townes \textit{et al.}\cite{Townes_PhysRev.76.1415} found clear disagreements with the spherical nuclear shell model,\cite{Mayer_PhysRev.75.1969,Jensen_PhysRev.75.1766.2} in particular in regions far away from closed shells.  Reconciliation required the development of theories of nuclear deformation.\cite{Rainwater_RevModPhys.48.385,Bohr_RevModPhys.48.365,Mottelson_RevModPhys.48.375,Moszkowski_HP1957, Ring_Schuck} These refinements also introduce a distinction between the intrinsic nuclear electric quadrupole moment $Q_0$, reflecting the nuclear deformation,\cite{Thyssen_PhysRevA.63.022505} and the spectroscopic nuclear electric quadrupole moment $Q$, observed in the laboratory frame. 

 Tables of (spectroscopic) nuclear electric quadrupole moments (NQMs) have been provided at regular intervals by Pyykkö \cite{Pyykko_MP2001,Pyykko_MP2008,Pyykko_MP2018}, likewise by Stone.\cite{Stone_ADNDT2005,Stone_ADNDT2016} The latest compilation by Stone\cite{Stone_NQM2021} lists for each isotope the method used for the determination of the nuclear electric quadrupole moment, and therefore allows to make statistics on the many experimental methods employed. If we consider all isotopes, then the dominant experimental methods are collinear laser spectroscopy\cite{Koszorus_EJP2024,Kaufman_OC1976} (23.5\%), Coulomb excitation reorientation\cite{Kean_qCE1976} (12.9\%) and time-dependent perturbed angular correlation\cite{Neugart_Neyens_2006} (9.9\%). However, these are methods mostly used for short-lived species.\cite{Neugart_Neyens_2006} If we restrict attention to stable isotopes, then the dominant methods are atomic beam magnetic resonance (32.1\%), muonic X-ray hyperfine structure\cite{Antognini_PhysRevC.101.054313} (25.6\%) and molecular spectroscopy (12.8 \%).

The basic interaction is the coupling of the NQM with the electric field gradient generated by electrons and other nuclei in the vicinity of the target nucleus. A direct determination of the nuclear electric quadrupole moment $eQ$ is possible by combining nuclear quadrupole coupling constants (NQCC) $e^{2}qQ/h$ obtained from experiment with accurate calculations of the electric field gradient (EFG) $eq$
\begin{equation}
    \text{NQCC [in MHz]}=234.9647\,\times\,Q\text{[in b]}\times q\,\left[\text{in }E_{h}/a_{0}^{2}\right]
\end{equation}
(see, for instance, Ref.~\citenum{Wang_JCP2013}). Once such a value is known, NQMs of other isotopes of the same element may be obtained from ratios of NQCCs, under the assumption of an ideal  point-like nucleus.\cite{Koszorus_EJP2024} 

 The electric field gradient is highly sensitive to small changes in the charge distribution around the nucleus which can occur, for example, in the formation of a chemical bond. It can therefore give us chemical information, as pointed out by Dailey and Townes.\cite{Dailey_JCP1949} In a 1949 paper, focusing on halogen compounds, these authors proposed  that the electric field gradient at a given nuclear position is determined by the partially occupied valence $p$ orbitals of the same atom. This in turn suggested that information about these populations could be inferred from knowledge of atomic and molecular NQCCs.\cite{Dailey_JCP1955} Subsequent refinements have been proposed in the literature.\cite{Cotton_Harris_PNAS1966,Yu1995,Nikitin_RCB1995,Novick_JMS2011,Rinaldi2020} A detailed study of the chemical information associated with the electric field gradient has been reported by Autschbach and co-workers.\cite{Autschbach2010}

In the present contribution we show that the Dailey--Townes model can be formulated as a particular approximation to projection analysis,\cite{Dubillard_JCP2006} providing a decomposition of expectation values in terms of atomic reference orbitals. This allows a detailed study of the validity of the Dailey--Townes model. Our exploration begins with a comprehensive examination of compounds featuring \ce{Cl} in diverse molecular systems, as explored in the seminal work of Dailey and Townes.\cite{Dailey_JCP1949} These include \ce{Cl2}, \ce{ICl}, \ce{ClCN}, \ce{CH3Cl}, and \ce{NaCl}. Subsequently, we extend our investigation to dihalogens of chlorine, \ce{X-Cl}, where \ce{X}=\ce{I}, \ce{At}, \ce{Ts}. Finally, we shall consider a molecule where bonding is dominated by $d$- and $f$-orbitals, namely uranyl, which also features a subvalence $6p$-hole, contributing significantly to the EFG.

Our paper is organized as follows: In Section \ref{sec:theory} we provide the basic theory of nuclear quadrupole coupling and present the Dailey--Townes model in the framework of projection analysis. In Section \ref{sec:compdet} we detail our computational protocol. The results of our analysis are presented and discussed in Section \ref{sec:results}, before concluding and providing perspective in Section \ref{sec:concl}.

Theory is presented in full SI units, whereas results are reported in SI-based atomic units.\cite{SIau}

\section{Theory}\label{sec:theory}
\subsection{Nuclear quadrupole coupling}
The electric quadrupole moment of some nucleus $K$ couples to the electric field gradient at the nuclear position, that we for convenience locate at the origin. The interaction energy is
\begin{equation}\label{eq:CartesianE2}
    E^K_{E2}=-\frac{1}{2}\sum_{\alpha\beta}Q_{\alpha\beta}^{[2]}\mathcal{E}_{\alpha;\beta}^{\left[1\right]};\quad Q_{\alpha\beta}^{[2]}=\int r_{\alpha}r_{\beta}\rho^K_{n}\left(\boldsymbol{r}\right)d^{3}\mathbf{r}
\end{equation}
(see for instance Ref.\citenum{TSAUEBOOK}). Components $\mathcal{E}_{\alpha;\beta}^{\left[1\right]}$ of the electric field gradient are generated by the charge densities associated with electrons and other nuclei
\begin{equation}
    \rho(\mathbf{r})=\rho_e(\mathbf{r})+\sum_{A\ne K}\rho^A_n(\mathbf{r})
\end{equation}
and can therefore be expressed as
\begin{equation}
    \mathcal{E}_{\alpha;\beta}^{\left[1\right]}=\left.\frac{\partial \mathcal{E}_{\alpha}}{\partial r_{\beta}}\right|_{\boldsymbol{r}=\boldsymbol{0}}=
    -\left.\frac{\partial^2\phi}{\partial r_{\alpha}\partial r_{\beta}}\right|_{\boldsymbol{r}=\boldsymbol{0}}=
    -\chi_{\alpha\beta}+\frac{1}{3\varepsilon_0}\delta_{\alpha\beta}\rho_e(\boldsymbol{0}),
\end{equation}
where appears the \textit{traceless} part of the \textit{negative} electric field gradient
\begin{equation}
    \chi_{\alpha\beta}=\frac{1}{4\pi\varepsilon_0}\int\rho(\mathbf{r})\frac{3r_{\alpha}r_{\beta}-\delta_{\alpha\beta}r^2}{r^5}d^3\mathbf{r}.
\end{equation}

The above interaction, Eq.\eqref{eq:CartesianE2}, is expressed in terms of components $Q_{\alpha\beta}^{[2]}$ of the \textit{Cartesian} electric quadrupole moment tensor. In terms of the \textit{traceless} electric quadrupole moment tensor\cite{Buckingham_QR1955}
\begin{equation}
    \Theta_{\alpha\beta}^{\left[2\right]}=\frac{3}{2}\left(Q_{\alpha\beta}^{[2]}-\delta_{\alpha\beta}\frac{1}{3}\sum_{\gamma}Q_{\gamma\gamma}^{\left[2\right]}\right)
\end{equation}
the interaction reads
\begin{equation}\label{eq:tracelessE2}
    E^K_{E2}=\frac{1}{3}\sum_{\alpha\beta}\Theta_{\alpha\beta}^{[2]}\chi_{\alpha\beta}-\frac{1}{6}\sum_{\alpha}Q_{\alpha\alpha}^{\left[2\right]}\sum_{\beta}\mathcal{E}_{\beta;\beta}^{\left[1\right]}
\end{equation}
The second term, denoted the Poisson term, contains the divergence of the electric field at the nuclear position
\begin{equation}
    \sum_{\beta}\mathcal{E}_{\beta;\beta}^{\left[1\right]}=\left.\left(\boldsymbol{\nabla}\cdot\boldsymbol{\mathcal{E}}\right)\right|_{\boldsymbol{r}=\boldsymbol{0}}=-\left.\nabla^{2}\phi\left(\boldsymbol{r}\right)\right|_{\boldsymbol{r}=\boldsymbol{0}}=\rho_{e}\left(\boldsymbol{0}\right)/\varepsilon_{0}\ne0,
\end{equation}
which by Gauss's law equals the source density at the expansion point divided by the electric constant $\varepsilon_{0}$. In many cases the source density is zero at the expansion point, such that the quadrupole interaction is fully described in terms of the traceless form, hence showing its utility. For nuclear quadrupole coupling, though, this is not the case, since the electronic charge density $\rho_e$ can not be expected to be zero at the nuclear origin. However, since the Poisson term is a product of two traces, it is perfectly isotropic and can in practice be dropped since it does not contribute to hyperfine splittings.

This can be seen explicitly by noting that, using angular momentum algebra, the components of the traceless nuclear electric quadrupole moment for a nuclear spin state $|I, m_I\rangle$ can be expressed in terms of nuclear spin operators as\cite{TSAUEBOOK}
\begin{equation}
    \Theta_{\alpha\beta}^{[2]}=\frac{eQ}{\hbar^2 I(2I-1)}\langle I,m_I|\frac{1}{2}(3\hat{I}_{\alpha}\hat{I}_{\beta}-\delta_{\alpha\beta}\hat{I}^2)|I,m_I\rangle .
\end{equation}
In the above expression appears the electric quadrupole moment $Q$ of the nucleus, defined by $eQ=2\Theta_{zz}$ where $e$ is the fundamental charge. By similar arguments one finds that
\begin{equation}
    \sum_{\alpha}Q_{\alpha\alpha}^{\left[2\right]}=\frac{eQ}{\hbar^2 I(2I-1)}\langle I,m_I|\hat{I}^2|I,m_I\rangle=\frac{eq(I+1)}{(2I-1)}.
\end{equation}
The interaction energy associated with a specific nuclear spin level can thereby 
be expressed as
\begin{equation}
    E_{E2}(I,m_I)=\frac{eQ}{\hbar^2 6I(2I-1)}\sum_{\alpha\beta}\langle I,m_I|(3\hat{I}_{\alpha}\hat{I}_{\beta}-\delta_{\alpha\beta}\hat{I}^2)|I,m_I\rangle\chi_{\alpha\beta}-\frac{1}{6\varepsilon_{0}}\frac{eQ(I+1)}{(2I-1)}\rho_{e}\left(\boldsymbol{0}\right)
\end{equation}
where the latter term is independent of the azimuthal quantum number $m_I$, hence do not contribute to hyperfine splittings. It should be noted that the above expression ignores electron-nucleus overlap effects.\cite{Thyssen_PhysRevA.63.022505,Koch_PhysRevA.81.032507} Furthermore, Sternheimer pointed out that the presence of a nuclear electric quadrupole moment should induce a quadrupole in the electron cloud, for instance the atomic core, which were thought to otherwise not contribute to the $E2$ interaction.\cite {Sternheimer_PhysRev.80.102.2} At first sight the resulting Sternheimer shielding appears to be not included in our calculations, but this is not the case, as explained in Appendix \ref{app:Sternheim}.

In the present work we focus on atoms and linear systems. Dropping the Poisson term, the interaction energy is given by
\begin{equation}\label{eq:E2b}
    E_{E2}\left(I,m_{I}\right)=\frac{e^{2}qQ}{4I\left(2I-1\right)}\left[3m_{I}^{2}-I\left(I+1\right)\right]
\end{equation}
The electric field gradient (EFG) $q$ is defined through $eq=\chi_{zz}$, that is, the traceless part of the $zz$-derivative of the scalar potential at the nuclear position. From axial symmetry it follows that $\chi_{xx}=\chi_{yy}=-\frac{1}{2}\chi_{zz}$.

In the present work we focus on the chemical information that can be obtained from knowledge of the electric field gradient at nuclear centers in a molecule, notably as suggested by the Dailey--Townes model. It turns out that this model can be considered a particular approximation to the decomposition of expectation values by projection analysis,\cite{Dubillard_JCP2006,bast:pncana_PCCP2011} as will be shown in the following.

\subsection{The Dailey--Townes model}\label{sec:DT}
The electric field gradient $q$ at the position of nucleus $K$ in an $N$-electron molecule can be obtained as an expectation value
\begin{equation}\label{eq:EFGexp}
    \langle e\hat{q}_{K}\rangle=\underbrace{\left\langle \sum_{i=1}^{N}\frac{-e}{4\pi\varepsilon_{0}}\left[\frac{3z_{iK}^{2}-r_{iK}^{2}}{r_{iK}^{5}}\right]\right\rangle }_{\text{Electronic: }eq_{K}^{e}}+\underbrace{\sum_{A\ne K}\frac{Z_{A}e}{4\pi\varepsilon_{0}}\left[\frac{3Z_{AK}^{2}-R_{AK}^{2}}{R_{AK}^{5}}\right]}_{\text{Nuclear: }eq_{K}^{n}},
\end{equation}
where the nuclear contribution $eq_{K}^{n}$ (not considering rovibrational degrees of freedom\cite{Lucken_ch1972,Zeil_ch1972,Lucken_ch1983}) is a simple scalar. At the self-consistent field (SCF) level, the electronic contribution is
\begin{equation}
\label{eq:elexp}
    \langle e\hat{q}_{K}^{e}\rangle  =  \sum_{i}^{N}\langle\psi_{i}^{MO}|e\hat{q}_{K}^{e}|\psi_{i}^{MO}\rangle
\end{equation}
Projection analysis\cite{saue:smalium,Dubillard_JCP2006} expresses molecular orbitals as a linear combination of atomic orbitals (LCAO)
\begin{equation}\label{eq:lcao}
    |\psi_{i}^{MO}\rangle=\sum_{A}\sum_{p\in A}c_{p,i}^{A}|\psi_{p}^{A}\rangle+|\psi_{i}^{\text{pol}}\rangle
\end{equation}
very much in line with common chemical thinking.\cite{Lennard-Jones_TF1929,Pauling_CR1928} It should be stressed that these are pre-calculated orbitals of the atoms constituting the molecule. By default, the expansion is limited to atomic orbitals occupied in the atomic ground state. The orthogonal complement $\psi_{i}^{\text{pol}}$ is denoted the polarization contribution. Significant polarization complicates analysis and may be a sign that important contributions from other AOs are missing. The polarization contribution can be completely eliminated by transformation to intrinsic atomic orbitals (IAOs).\cite{Knizia_JCTC2013} However, in the present setting the polarization contribution carries its name aptly since it provides a precise definition of the difference between the molecular space and the superposition of the free atomic ones. 
Inserting the expansion Eq.~\eqref{eq:lcao} into the above expectation value, Eq.~\eqref{eq:elexp} provides a decomposition into intra-atomic and inter-atomic contributions, as well as contributions associated with polarization
\begin{equation}
\label{eqn:prjanalysis}
    \langle e\hat{q}_{K}^{e}\rangle = \sum_{A}\langle e\hat{q}_{K}^{e}\rangle_{\text{intra}}^{\left(A\right)}+\sum_{A}\sum_{B\ne A}\langle e\hat{q}_{K}^{e}\rangle_{\text{inter}}^{\left(A,B\right)}+\langle e\hat{q}_{K}^{e}\rangle_{\text{pol}}.
\end{equation}

In passing, we note that if $e\hat{q}_{K}^{e}$ above is replaced by the identity operator, we obtain a decomposition of the integrated density (electrons) in line with Mulliken population analysis,\cite{mulliken:pop} with net populations given by intra-atomic contributions and gross populations defined in the usual way as
\begin{equation}\label{eq:popana}
g(A)=\langle 1\rangle_{\text{intra}}^{\left(A\right)}+\frac{1}{2}\sum_{A}\sum_{B\ne A}\langle 1\rangle_{\text{inter}}^{\left(A,B\right)}.
\end{equation}
Further decomposition in terms of atomic orbitals gives effective electronic configurations of atoms within the molecule. A crucial difference with respect to Mulliken population analysis, though, is that projection analysis is very stable with respect to a change of basis.\cite{Dubillard_JCP2006}

If we now focus on the intra-atomic contribution $\langle e\hat{q}_{K}^{e}\rangle_{\text{intra}}^{\left(K\right)}$ to $eq_{K}^{e}$ from the same center, it can be written as
\begin{equation}\label{eq:intraK}
    \langle e\hat{q}_{K}^{e}\rangle_{\text{intra}}^{\left(K\right)}=\underbrace{\sum_{p\in K}D_{pp}^{KK}\langle\psi_{p}^{K}|e\hat{q}_{K}^{e}|\psi_{p}^{K}\rangle}_{\text{principal}}+\underbrace{\sum_{p\in K}\sum_{\left(q\ne p\right)\in K}D_{pq}^{KK}\langle\psi_{q}^{K}|e\hat{q}_{K}^{e}|\psi_{p}^{K}\rangle}_{\text{hybridization}},
\end{equation}
where appears a density matrix
\begin{equation}
    D_{pq}^{AB}=\sum_{i}^{N}c_{pi}^{A}c_{qi}^{B\ast},
\end{equation}
built from the expansion coefficients of Eq.~\eqref{eq:lcao}. We distinguish between principal contributions, which would also contribute to the atomic expectation value, and hybridization contributions, which arises due to the mixing of orbitals of a given atom within the molecule.\cite{coulson:valence}

Since the matrix elements of $\langle e\hat{q}_{K}^{e}\rangle_{\text{intra}}^{\left(K\right)}$ are limited to atomic orbitals of the same center, much information is provided by taking into account the structure of the atomic orbitals. Throughout this work we shall mostly limit attention to the case where spin-orbit interaction is not included, in line with the Dailey--Townes model. The general form of scalar-relativistic, as well as non-relativistic, atomic orbitals is
\begin{equation}\label{eq:scalar AOs}
    \psi_{n\ell m}=R_{n\ell}\left(r\right)Y_{\ell m}\left(\theta,\phi\right).
\end{equation}
Placing the nucleus $K$ at the origin, the electronic EFG operator can be expressed as
\begin{equation}
    \frac{3z^{2}-r^{2}}{r^{5}}=\frac{3\cos^{2}\theta-1}{r^{3}}=\frac{2}{r^3}C^2_0\left(\theta,\phi\right),
\end{equation}
where appears a component of the spherical harmonic tensor\cite{Casimir_book1936,Racah_PhysRev.62.438}
\begin{equation}\label{eq:sphtens}
C^{\ell}_{m}\left(\theta,\phi\right)=\left(\frac{4\pi}{2\ell+1}\right)^{1/2}Y_{\ell m}\left(\theta,\phi\right).
\end{equation}
Matrix elements can then be factorized into radial and angular parts
\begin{equation}\label{eq:nrEFG}
    \langle\psi_{n\ell m}^{K}|e\hat{q}_{K}^{e}|\psi_{n^{\prime}\ell^{\prime}m^{\prime}}^{K}\rangle_{r,\theta,\phi}=-\frac{2e}{4\pi\varepsilon_{0}}\langle R_{n\ell}|r^{-3}|R_{n^{\prime}\ell^{\prime}}\rangle_{r}\langle Y_{\ell m}|C^{2}_{0}|Y_{\ell^{\prime}m^{\prime}}\rangle_{\theta,\phi}.
\end{equation}
From consideration of angular momentum and parity we obtain the following selection rules
\begin{equation}
    \ell^{\prime}\in\left\{ \left|\ell-2\right|,\ell,\ell+2\right\} \quad\land\quad m^{\prime}=m.
\end{equation}
Furthermore, within an atomic shell ($\ell^{\prime}=\ell$) matrix elements are given by
\begin{equation}\label{eq:atomic shell}
    \langle\psi_{n\ell m}^{K}|e\hat{q}_{K}^{e}|\psi_{n\ell m^{\prime}}^{K}\rangle_{r,\theta,\phi}=-\frac{2e}{4\pi\varepsilon_{0}}\langle r^{-3}\rangle\frac{\ell\left(\ell+1\right)-3m^{2}}{\left(2\ell-1\right)\left(2\ell+3\right)}\delta_{mm^{\prime}}.
\end{equation}
We note in particular the appearance of the inverse cubic radial expectation value $\langle r^{-3}\rangle$, which shows that the EFG probes the core region around the nucleus, making it sensitive to relativistic effects.

The chemical interest of the EFG arises from Uns\"olds theorem, \cite{UNSOLD} which essentially states that a closed atomic shell is spherically symmetric. Its contribution to the EFG expectation value is thereby zero, that is,
\begin{equation}
    \sum_{m=-\ell}^{\ell}\left|Y_{\ell m}\left(\theta,\phi\right)\right|^{2}=\frac{2\ell+1}{4\pi}\quad\Rightarrow\quad\sum_{m=-\ell}^{\ell}\langle Y_{\ell m}|C^{2}_{0}|Y_{\ell m}\rangle_{\theta,\phi}=0.
\end{equation}
(We note in passing that the left-hand side of the above equations shows the utility of the normalization chosen for the spherical harmonic tensors, Eq.~\eqref{eq:sphtens}.) This observation suggests that core orbitals do not contribute to the EFG expectation value. Instead, the operator $e\hat{q}_{K}^{e}$ samples the inner tails of valence orbitals, making it sensitive to electron correlation. The need to properly account for both relativity and electron correlation makes the accurate calculation of the EFG quite challenging, as illustrated by the determination of the NQM of \ce{^{197}Au}\cite{Belpassi_JCP2007,Thierfelder_PhysRevA.76.034502} and \ce{^{209}Bi}.\cite{Bieron_PhysRevLett.87.133003,Teodoro_PhysRevA.88.052504,Shee_JCP2016,Pyykko_MP2018,Skripnikov_PhysRevC.104.034316,Dognon_PCCP2023,Joosten_PhysRevB.110.045141} An exemplary recent recent calculation has been reported by Skripnikov and Barzakh.\cite{Skripnikov_PhysRevC.109.024315}

The basis of the Dailey--Townes (DT) model is the above observation that the EFG is sensitive to deviations from spherical symmetry. Such deviations can be induced by chemical bonding. Since core orbitals as well as valence $s$-orbitals are not expected to contribute and that positive and negative charge on other centers possibly cancel out (\textit{vide infra}), Dailey and Townes suggested that the EFG is entirely due to the contribution of valence $p$-orbitals on the atom of interest. In the present notation their approximation therefore reads
\begin{equation}\label{eq:DTmodel}
    \langle e\hat{q}_{K}\rangle\approx\langle e\hat{q}_{K}^{e}\rangle_{\text{intra}}^{\left(K\right)}\approx \frac{2e}{4\pi\varepsilon_{0}}\langle r^{-3}\rangle_{np}\times\frac{2}{5}\left(n_x+n_y-2n_z\right);\quad n_{\alpha}=D^{KK}_{np_{\alpha};np_{\alpha}},
\end{equation}
where we used Eq.~\eqref{eq:atomic shell}. 

The angular part of Eq.~\eqref{eq:nrEFG} may be evaluated using angular momentum algebra, notably the Wigner--Eckart theorem.\cite{lindgren_morrison} The same tools may be used when extending the formalism to include spin-orbit interaction.\cite{Grant:book} We start from the general form of 4-component atomic orbitals
\begin{equation}\label{eq:4c orbital}
\psi_{njm_j}=\left[\begin{array}{c}
R_{n\kappa}^{L}\left(r\right)\xi_{\kappa,m_{j}}\left(\theta,\phi\right)\\
iR_{n\kappa}^{S}\left(r\right)\xi_{-\kappa,m_{j}}\left(\theta,\phi\right)
\end{array}\right]
\end{equation}
with real scalar radial functions for the large (L) and small (S) components and 2-component complex angular functions. The angular functions are defined in terms of quantum numbers $j$,$m_j$,$\ell$ and $\kappa$, with the sign of $\kappa=2(\ell-j)(j+1/2)$ indicating from which orbital angular momentum $\ell$ the total angular momentum $j$ is generated.\cite{Grant:book,Szmytkowski2007} Intra-atomic matrix elements over the EFG operator are given by
\begin{eqnarray}
    \langle\psi^K_{njm_j}|e\hat{q}_{K}^{e}|\psi^K_{n^{\prime}j^{\prime}m^{\prime}_{j}}\rangle_{r,\theta,\phi}
    &=&
    -\frac{2e}{4\pi\varepsilon_{0}}\langle R^L_{n\kappa}|r^{-3}|R^L_{n^{\prime}\kappa^{\prime}}\rangle_{r}\langle \xi_{\kappa, m_j}|C^{2}_{0}|\xi_{\kappa^{\prime},m^{\prime}_{j}}\rangle_{\theta,\phi}\\
    &&
    -\frac{2e}{4\pi\varepsilon_{0}}\langle R^S_{n\kappa}|r^{-3}|R^S_{n^{\prime},\kappa^{\prime}}\rangle_{r}\langle \xi_{-\kappa,m_j}|C^{2}_{0}|\xi_{-\kappa^{\prime}m^{\prime}_{j}}\rangle_{\theta,\phi}\nonumber
\end{eqnarray}
However, complete factorization into radial and angular parts is possible since the angular integrals are independent of the sign of $\kappa$.\cite{Grant:book}
Again, from consideration of angular momentum and parity, we find that $p_{3/2;m_j}$ couples to $p_{1/2;m_j}$, $p_{3/2;m_j}$, $f_{5/2;m_j}$ and $f_{7/2;m_j}$. Likewise, $p_{1/2;m_j}$ couples to $p_{3/2;m_j}$ and $f_{5/2;m_j}$, but not itself. The latter follows from the same argument as used above to conclude that nuclear spin $I>1/2$ is required to have a nuclear electric quadrupole moment; a $p_{1/2}$-orbital, just like $s_{1/2}$, has a spherical density. 

We shall again be specifically interested in matrix elements within an atomic shell characterized by (large component) orbital angular momentum $\ell$. Possible values of total angular momentum are then $j=\ell+1/2\eta$, where   $\eta=\pm1$. Matrix elements are given by
\begin{equation}\label{eq:matrix_SO}
    \langle\psi^K_{n,\ell+1/2\eta_{1},m_j}|e\hat{q}_{K}^{e}|\psi^K_{n\ell+1/2\eta_{2},m^{\prime}_{j}}\rangle_{r,\theta,\phi}=
    -\frac{2e}{4\pi\varepsilon_{0}}\langle r^{-3}\rangle_{\eta_1,\eta_2}\delta_{m_j,m^{\prime}_{j}}A_{\eta_1,\eta_2;m_j}.
\end{equation}
It should be noted that with the introduction of spin-orbit interaction, the operator  $e\hat{q}_{K}^{e}$ is no longer diagonal within an atomic shell of given orbital angular momentum $\ell$, since there is now generally coupling between the SO-components of $j=\ell\pm 1/2$.
Radial factors are given by
\begin{equation}
    \langle r^{-3}\rangle_{\eta_1,\eta_2}=
    \langle R^L_{n\kappa_1}|r^{-3}|R^L_{n\kappa_2}\rangle_{r}
    +\langle R^S_{n\kappa_1}|r^{-3}|R^S_{n\kappa_2}\rangle_{r};\quad\kappa_i=-\eta_i\left[\ell+\frac{1}{2}(\eta_i+1)\right],
\end{equation}
whereas angular factors are given by
\begin{equation}
    A_{\eta_1,\eta_2;m_j}=\langle\ell+1/2\eta_1,m_j|C_{0}^{\left(2\right)}|\ell+1/2\eta_2,m_j\rangle_{\theta,\phi}.
\end{equation}
Specifically, one has
\begin{equation}
\begin{array}{lcrcr}
A_{--;m_j} & = & \displaystyle\frac{\left(2\ell+1\right)\left(2\ell-1\right)-12m_j^{2}}{4\left(2\ell+1\right)\left(2\ell-1\right)};&\quad &2\ell\ge3\\\\
A_{+-;m_j}& = & \displaystyle -\frac{3m_j\sqrt{\left(2\ell+1\right)^2-4m^2_j}}{\left(2\ell+1\right)\left(2\ell-1\right)\left(2\ell+3\right)};&\quad&\ell\ge1\\\\
A_{++;m_j} & = & \displaystyle\frac{\left(2\ell+1\right)\left(2\ell+3\right)-12m_j^{2}}{4\left(2\ell+1\right)\left(2\ell+3\right)};&\quad&2\ell\ge1
\end{array}\label{eq:angfac}
\end{equation}
In passing, we note that the sign of the mixed angular factor $A_{+-;m_j}$ depends on the sign of $m_j$. Under time reversal, the two-component angular functions $\xi_{\kappa,m_j}$ transform as
\begin{equation}
    {\cal K}\xi_{\kappa,m_j}=\mbox{sgn}(\kappa)(-1)^{m_j+1/2}\xi_{\kappa,-m_j};\quad {\cal K}=-i\sigma_y{\cal K}_0,
\end{equation}
where ${\cal K}_0$ is complex conjugation. This implies that
\begin{eqnarray}
    {\cal K}\langle\ell+1/2,m_j|C_{0}^{\left(2\right)}|\ell-1/2,m_j\rangle_{\theta,\phi}
    &=&-\langle\ell+1/2,-m_j|C_{0}^{\left(2\right)}|\ell-1/2,-m_j\rangle_{\theta,\phi}\\
    &=&\langle\ell+1/2,m_j|C_{0}^{\left(2\right)}|\ell-1/2,m_j\rangle_{\theta,\phi},
\end{eqnarray}
which is the expected behaviour from a time-symmetric operator.
Casimir does not explicitly give the above matrix elements.\cite{Casimir_book1936} Rather, he considers the expectation value of $e\hat{q}_{K}^{e}$ for states $|J,M_J=J\rangle$ arising from $s,\ell$-configurations, e.g the excited \ce{^3L} (5s5d) state of indium, starting either from $LS$-coupling (\textsection 9),  or from $jj$-coupling (\textsection 15). In the latter case his expressions are consistent with the above matrix elements.

In the following we will explore the Dailey--Townes model using the projection analysis and decomposition of SCF expectation values implemented in the DIRAC code for relativistic molecular calculations.\cite{DIRACpaper2020} We first consider the same halogen molecules, dominated by valence $p$ bonding, as studied by Dailey and Townes in their original publication. We will then investigate how spin-orbit interaction affects the analysis. Finally we will consider the uranyl molecule, where bonding is mediated by uranium $d$ and $f$ orbitals.

\section{Computational details}\label{sec:compdet}
We performed all calculations with the DIRAC program for relativistic molecular calculations,\cite{DIRACpaper2020,DIRAC24} employing a value of 137.035 999 8 $a_0 E_h/\hbar$ for the speed of light. A Gaussian model for the nuclear charge distribution was  employed throughout our calculations, using the parameters of Ref.~\citenum{Visscher:atomic}. We have carried out relativistic Kohn--Sham calculations\cite{DFT_DIRAC2,DFT_DIRAC3} based on the 4-component Dirac-Coulomb Hamiltonian, using the GGA (Generalized Gradient Approximation) exchange-correlation functional PBE.\cite{PBE} Unless otherwise stated, and in order to connect to the Dailey--Townes model, the calculations included only scalar relativistic effects, the spin-orbit interaction eliminated as described in Ref.~\citenum{saue:spinfree}. All the halogens were equipped with a dyall.3zp basis set,\cite{dyall:4dbasis} suitable for SCF calculations. For \ce{UO2^2+}, we have used the slightly larger dyall.v3z basis set for all atoms.\cite{Dyall2006,Dyall2016} 
%\ce{UO2^2+} has a group symmetry $D_{\infty h}$ but we have reduced the symmetry to $C_{\infty h}$ introducing a ghost center. In this way oxygen atoms are no longer connected by symmetry and we can carry out the projection analysis in $C_{\infty h}$.  Since uranyl is a polyatomic molecule, the MOs that we get after the SCF procedure are delocalized. 
Molecular orbitals were localized using the Pipek-Mezey criterion\cite{Pipek-Mezey_JCP1989} combined with an exponential parametrization and a trust region minimization method.  
For the dihalogen compounds, we have calculated and used the bond distances reported in Table \ref{tab:bondX}. The bond distance 1.7044 {\AA} of \ce{UO2^2+} was taken from Ref.\citenum{bondUO2} .

\begin{table}[H]
    \centering
    \begin{tabular}{l | c c c c c c}
    & \ce{FCl} &\ce{Cl2} & \ce{BrCl} & \ce{ICl} & \ce{AtCl} &\ce{TsCl}\\
    \hline
    \hline
         no-SO& 1.664&2.043&2.186& 2.322& 2.419&2.522\\
         SO& 1.665&2.026&2.182& 2.374&2.523&2.712\\
    \end{tabular}
    \caption{Optimized bond distances (in {\AA}) for chlorine dihalogens. The distances were calculated using 4-component Dirac-Coulomb Hamiltonian with DFT/PBE, with and without spin-orbit (SO) interaction.}
    \label{tab:bondX}
\end{table}

\section{Results and discussion}\label{sec:results}
%In this section, the key findings of this study will be presented. The focus lies on the computed electric field gradient values at the positions of \ce{^35Cl} in \ce{X\bond{-}Cl} type molecules (X=F, Cl, Br, I, At, and Ts) and \ce{^{235}U} in \ce{UO2^{2+}}. The analysis begins by scrutinizing the electric field gradient for the bare \ce{^35Cl} nucleus, ensuring the alignment of results with the Dailey--Townes model. Subsequently, the investigation extends to dihalogen compounds (\ce{X\bond{-}Cl}, where X=F, Cl, Br, I, At, and Ts) to explore variations in the electric field gradient at the \ce{^35Cl} position induced by changes in the bonded atom. This exploration includes an examination of the impact of spin-orbit coupling on the accurate calculation of the electric field gradient for these compounds. A more intricate analysis is reserved for the molecule ICl, where a comprehensive evaluation of the electric field gradient will be conducted. Through projection analysis, the atomic orbitals contributing significantly to the electric field gradient will be identified. Finally, the electric field gradient at the position of \ce{^{235}U} in the \ce{UO2^{2+}} molecule will be scrutinized. This examination aims to extract chemical insights into the U-O chemical bond.

\subsection{Chlorine compounds}
In this section we will examine the chemical information that can be obtained from the electric field gradient in some chlorine compounds, most of which are present in the original paper of Dailey and Townes\cite{Dailey_JCP1949} (\ce{ICl}, \ce{ClCN}, \ce{CH3Cl} and \ce{NaCl}), but also including \ce{Cl2}. We start by considering the chlorine atom itself. In Table \ref{tab:tabpol} we present the matrix elements $\braket{e\hat{q}_{Cl}}_{ii}$ of individual orbitals obtained from a PBE calculation using fractional occupation corresponding to the ground state configuration [Ne]3s$^{2}$3p$^{5}$; this occupation is reported as Case I in the Table. We first note that the values of the matrix elements are consistent with Eq.~\eqref{eq:atomic shell}, giving inverse cubic radial expectation values $\langle r^{-3}\rangle$ of 101.98 $a_0^{-3}$ and 7.512 $a_0^{-3}$, respectively, for the 2p and 3p shells. We also observe that summing expectation values within an atomic shell indeed gives zero in accordance with Uns\"old's theorem.

\begin{table}[H]
    \centering
    \begin{tabular}{ c c c c}
         Orbital & $\braket{e\hat{q}_{Cl}}_{ii}$ & Case I& Case II\\
         \hline
         1s&  0.000&2.000& 2.000\\ 
         \cline{2-4}
         2s&   0.000&2.000&2.000\\
         \cline{2-4}
         2p$_{x}$&  40.792& 2.000& 2.000\\
         2p$_{y}$&   40.792& 2.000&2.000\\
         2p$_{z}$&  -81.584& 2.000& 2.000\\
         \cline{2-4}
          Total&  &0.000 &0.000 \\
         \cline{2-4}
         3s& 0.000&2.000& 2.000\\
         \cline{2-4}
         3p$_{x}$& 3.005 &1.670&2.000\\
         3p$_{y}$& 3.005 &1.670&2.000\\
         3p$_{z}$& -6.010 &1.670&1.000\\
         \cline{2-4}
         Total &  &0.000& 6.010 \\
         \hline
    \end{tabular}
    \caption{\footnotesize Interatomic contributions $\braket{e\hat{q}_{Cl}}^{Cl}$ calculated from matrix elements and net occupations. Two sets of occupations are considered: Case I is the fractional occupation of the atomic ground state, whereas Case II is the idealized occupation in \ce{Cl2}. All values of $eq_{Cl}$ are in atomic units (E$_{h}/ea_{0}^{2}$). }
    \label{tab:tabpol}
\end{table}

We next turn to molecular species. In Table \ref{tab:we} we have listed the title molecules \ce{ClX} according to the Pauling electronegativity difference $\Delta\chi_{Cl-X}$. For the chlorine molecule one would ideally expect an occupation corresponding to Case II of Table \ref{tab:tabpol}. The sum of matrix elements $\braket{e\hat{q}_{Cl}}_{ii}$ weighted by this occupation gives the value
\begin{equation}\label{eq:Cl2}
    eq_{Cl}\left[\ce{Cl2}\right]=\frac{2e}{4\pi\varepsilon_{0}}\langle r^{-3}\rangle_{3p}\times\frac{4}{5}=6.010 E_{h}/ea_{0}^{2}
\end{equation}
which is remarkably close to the total value $eq_{Cl}^{SCF}$ = 6.075 E$_{h}/ea_{0}^{2}$ reported in Table \ref{tab:we}. However, the actual net populations of the chlorine 3p orbitals obtained in our calculations are somewhat different, due to overlap and hybridization. They are reported in Table \ref{tab:we}, giving a Dailey--Townes value $eq_{Cl}^{DT}$ of 7.345 E$_{h}/ea_{0}^{2}$. Generally, for the molecules in Table \ref{tab:we}, we expect that with increasing positive electronegativity difference $\Delta\chi_{Cl-X}$, leading to increased charge transfer towards the chlorine atom, the electric field gradient should decrease towards zero as we go to the limiting case of a chlorine anion. This is indeed what we see for the calculated total values $eq_{Cl}$, whereas the values $eq_{Cl}^{DT}$ obtained using the Dailey--Townes model of Eq.~\eqref{eq:DTmodel} are often in significant error and even unable to give the right trend.

\begin{table}[H]
\label{tab:tabb}
    \centering
    \begin{tabular}{l|c|c|c|c|c|c}
        Molecule & $\Delta\chi_{Cl-X}$& n$\left(3p_{x}\right)$ & n$\left(3p_{y}\right)$&n$\left(3p_{z}\right)$  &$eq_{Cl}^{DT}$& $eq_{Cl}$\\
        \hline
        \hline
         \ce{Cl2}    & 0.00 & 2.034 &2.034  &  0.812 & 7.345& 6.075 \\
                     &      & 1.992 &1.992  &  1.017 & 5.861&\\
         \ce{ICl}    & 0.50 & 2.030 &2.030  & 1.048  & 5.902& 4.736  \\
                     &      & 1.990 &1.990  & 1.240  & 4.507&\\
         \ce{ClCN}   & 0.61 & 1.902 & 1.902 & 0.855  &6.295 & 4.340 \\
                     &      & 1.897 & 1.897 & 1.154  &4.468 &\\
         \ce{CH3Cl}  & 0.61 & 2.049 & 2.056 & 1.036  &6.115 & 4.017 \\
                     &      & 1.978 & 1.978 & 1.240  & 4.436&\\
         \ce{NaCl}   & 2.23 & 1.976 & 1.976 & 1.830  &0.871 & 0.575 \\
                     &      & 1.975 & 1.975 & 1.873  &0.614\\
    \end{tabular}
  \caption{{ \footnotesize Calculated values (in E$_{h}/ea_{0}^{2}$) of the electric field gradient at the nuclear position \ce{Cl} of molecules \ce{ClX}. The molecules are listed according to the difference $\Delta\chi_{Cl-X}$ in electronegativity on the Pauling scale . The value $eq_{Cl}$ refers to the calculated total electric field gradient, including nuclear contributions, whereas $eq_{Cl}^{DT}$ is obtained with the Dailey--Townes model, that is, according to Eq.~\eqref{eq:DTmodel}. We also give the net populations on chlorine, obtained from projection analysis, used to calculate $eq_{Cl}^{DT}$. For each molecule, we give in the second line the numbers obtained when net populations are replaced by gross populations. These calculations were performed in the absence of spin-orbit interaction.}}
    \label{tab:we}
\end{table}

\begin{table}[h]
    \centering
    \begin{tabular}{l c c| c c |c c c c}
        Orb &$\varepsilon$&Sym &I & Cl &I(5$s$)& I(5$p$) &Cl(3$s$)&Cl(3$p$) \\
        \hline
        \hline
         29&-0.791 & $\sigma$&0.392 &1.598 & 0.320&0.071 & 1.554&0.044       \\ 
         30& -0.642 & $\sigma$& 1.560 &0.438 &1.553 & 0.006& 0.360&  0.078     \\ 
         31&  -0.395& $\sigma$&0.798 & 1.172 &0.091 & 0.708& 0.056& 1.117    \\
         32&-0.326 & $\pi$&0.558 &1.428 & 0.000& 0.558& 0.000& 1.429    \\
         33& -0.326&$\pi$ &0.558 &1.428& 0.000& 0.558& 0.000& 1.429   \\
         34& -0.243&$\pi$ &1.436 &0.560 & 0.000& 1.436& 0.000&0.560     \\
         35& -0.243&$\pi$ &1.436 &0.560 & 0.000& 1.436& 0.000&0.560      \\
    \end{tabular}
    \caption{Projection analysis of canonical molecular orbitals for \ce{ICl}. We report the gross populations for \ce{I} and \ce{Cl} (columns 4 and 5) as well as valence orbitals (columns 6-9). Orbital energies $\varepsilon$  are given in $E_{h}$.}
    \label{tab:ICl_canonical}
\end{table}

\begin{table}[h]
    \centering
    \begin{tabular}{l c c| c c |c c c c}
        Orb &$ \braket{\varepsilon}$&Sym &I & Cl &I(5$s$)& I(5$p$) &Cl(3$s$)&Cl(3$p$) \\
        \hline
        \hline
         29& -0.916&$\sigma$& 0.754&1.210 &0.005 &0.747 & 0.011& 1.199      \\ 
         30& -1.275&$\sigma$&2.020 &-0.022 & 1.963& 0.056& -0.003&       -0.016\\ 
         31&-1.463 &$\sigma$&-0.002&2.018 & -0.004& -0.017&1.963 &       0.056\\
         32& -0.533&$\pi$& 2.004&-0.004 & 0.000& 2.004& 0.000&     -0.009\\
         33& -0.533&$\pi$& 2.004&-0.004 & 0.000& 2.004& 0.000&    -0.009\\
         34& -0.605&$\pi$& -0.002& 1.998& 0.000& -0.009& 0.000&1.999     \\
         35& -0.605&$\pi$& -0.002& 1.998 & 0.000& -0.009& 0.000&1.999     \\
    \end{tabular}
    \caption{Projection analysis of localized molecular orbitals for \ce{ICl}. $ \braket{\varepsilon}$ refers to the expectation value (in $E_{h}$)  of the converged Kohn-Sham operator. We report the gross populations for \ce{I} and \ce{Cl} (columns 4 and 5) as well as valence orbitals (columns 6-9). }
    \label{tab:ICl_local}
\end{table}

In order to understand the limited success of the Dailey--Townes model, we shall single out the \ce{ICl} molecule for detailed analysis. This molecule was also scrutinized by Dailey and Townes, but within the framework of valence bond theory.\cite{Shaik_mol2021} They considered bonding to be mediated by a single $\sigma$-bond; starting from a Heitler--London covalent form they considered the effect of hybridization, overlap and ionicity. We, on the other hand, work within the setting of molecular orbital (MO) theory. Upon inspection of the canonical MOs (cf. Table \ref{tab:ICl_canonical}), we find, somewhat to our surprise, a more complicated picture, with seven MOs having significant gross population on both centers. However, upon localization, limited to the valence orbitals of Table \ref{tab:ICl_canonical}, we recover the bonding picture discussed by Dailey and Townes (cf. Table \ref{tab:ICl_local}). The now single $\sigma$ bonding orbital is almost exclusively spanned by $p_z$ orbitals on the two centers
\begin{equation}\label{eq:orb29}
    \psi_{29}\approx a \mbox{ Cl3p}_z + b \mbox{ I5p}_z,
    \quad\left\{\begin{array}{lcr}a&=&0.703\\b&=&-0.514\end{array}\right. .
\end{equation}
According to the Dailey--Townes model, Eq.~\eqref{eq:DTmodel}, we then get
\begin{equation}\label{eq:q_ICl}
    eq_{Cl}\left[\ce{ICl}\right]=\frac{2e}{4\pi\varepsilon_{0}}\langle r^{-3}\rangle_{3p}\times\frac{4}{5}(2-2a^2)=(2-2a^2)eq_{Cl}\left[\ce{Cl2}\right],
\end{equation}
where the right-hand side refers to the expression in Eq.~\eqref{eq:Cl2}. The \ce{ICl} bonding $\sigma$-orbital, Eq.~\eqref{eq:orb29}, to a very good approximation, involves only $p_z$ orbitals from the two centers. This is in vivid contrast to Dailey and Townes who suggested significant $s$ hybridization on the chlorine center, on the order of 20\%, in order to accommodate their model with experimental data.\cite{Dailey_JCP1949} A different point of view was taken by Gordy,\cite{Gordy_JCP1954,Gordy_FaradayDisc1955} who suggested that one could somehow ignore overlap when evaluating the electric field gradient. This would imply that the net populations appearing in the Dailey--Townes model, Eq.~\eqref{eq:DTmodel}, are replaced by gross populations. Indeed, if we proceed in this manner, we get much better agreement with the calculated total values $eq_{Cl}$, as shown in Table \ref{tab:we} (second row for each molecule). In the case of \ce{ICl} we could then combine the normalization $a^2+b^2=1$, which is clearly not true, with the definition of ionicity\cite{Gordy_FaradayDisc1955}
\begin{equation}
    \beta=a^2-b^2
\end{equation}
to rewrite Eq.~\eqref{eq:q_ICl} as
\begin{equation}
    \beta=1-\frac{eq_{Cl}\left[\ce{ICl}\right]}{eq_{Cl}\left[\ce{Cl2}\right]}
\end{equation}
For the left-hand side, using the coefficients from Eq.~\eqref{eq:orb29}, we get 0.230, whereas the right-hand side, using the calculated total values $eq_{Cl}$ from Table \ref{tab:we}, gives 0.220. Although the agreement is intriguing, we can see no obvious justification for the overlap neglect by Gordy. A better line of argument was provided by Cotton and Harris:\cite{Cotton_Harris_PNAS1966} They split (the valence) inter-atomic contributions $\langle e\hat{q}_{K}^{e}\rangle_{\text{inter}}^{\left(K,A\right)}$ in two equal parts. One part is combined with the inter-atomic contribution $\langle e\hat{q}_{K}^{e}\rangle_{\text{intra}}^{\left(A\right)}$ as well as the nuclear contribution from the same center $A$ and their sum assumed to be zero. For the second part, they assume a Mulliken-type relation\cite{Mulliken_1949a,Mulliken_1949b}
\begin{equation}
    \langle\psi^A_p|e\hat{q}_{K}^{e}\psi^K_q\rangle\approx\langle\psi^A_p|\psi^K_q\rangle\langle\psi^K_q|e\hat{q}_{K}^{e}|\psi^K_q\rangle .
\end{equation}
Inserted back into the expansion Eq.~\eqref{eqn:prjanalysis} one indeed gets a modified version of the Dailey--Townes model, now based on gross populations.

\begin{table}[H]
    \centering
    \begin{tabular}{c c| ccccccc}
           &  &$\braket{e\hat{q}^{\text{el}}_{\text{Cl}}}_{\text{princ}}^{(\text{Cl})}$&$\braket{e\hat{q}^{\text{el}}_{\text{Cl}}}_{\text{hyb}}^{(\text{Cl})}$ &$\braket{e\hat{q}^{\text{el}}_{\text{Cl}}}_{\text{intra}}^{(\text{Cl})}$  & $\braket{e\hat{q}^{\text{el}}_{\text{Cl}}}_{\text{intra}}^{(\text{I})}$  &$\braket{e\hat{q}^{\text{el}}_{\text{Cl}}}_{\text{inter}}^{(\text{Cl})}$ & $\braket{e\hat{q}^{\text{el}}_{\text{Cl}}}_{\text{pol}}^{(\text{Cl})}$  & $\braket{e\hat{q}^{\text{el}}_{\text{Cl}}}^{(\text{Cl})}$  \\
         \ce{I} core & &-0.000&-0.000&-0.000&-1.089 & -0.000&  -0.000& -1.091 \\
         \cline{1-9}
         \ce{Cl} core & & 0.000& 0.0455 & 0.045& 0.00 & 0.00 & 0.155  &0.2003  \\
         \cline{1-9}
           &$\sigma$ bond & -5.952 & -0.568 & -6.520 & -0.019& -0.026& -0.302  &-6.866  \\
           \rotatebox[origin=c]{90}{Valence} &$\sigma$ n.b & -0.362 & 0.013& -0.349 & -0.038& 0.008& -0.024& -0.403  \\
          &$\pi$ n.b& 12.205 & -0.060& 12.145 & -0.056& 0.001& -0.449  &11.641  \\
   &Total valence&5.891 &-0.616 &5.276 &-0.113 &-0.017&-0.775 & 4.372\\
    \end{tabular}
    \caption{Decomposition of electronic contribution to the EFG at the position of \ce{Cl} in \ce{ICl} using the projection analysis for each group of MOs. All values are reported in $\text{E}_{h}/ea_{0}^{2}$. Adding the nuclear contribution of 1.255 $\text{E}_{h}/ea_{0}^{2}$ gives the total contribution of 4.736 $\text{E}_{h}/ea_{0}^{2}$.  n.b=non-bonding. }
    \label{tab:ICl_decomp}
\end{table}

Let us now consider the complete decomposition of the EFG according to Eq.~\eqref{eqn:prjanalysis}. This is shown in Table \ref{tab:ICl_decomp}. We have divided the MOs into three groups: i) core orbitals on iodine, ii) core orbitals on chlorine and iii) valence orbitals, the latter corresponding to the MOs shown in Table \ref{tab:ICl_local}. For the valence orbitals we further distinguish between the bonding $\sigma$-orbital (MO 29), non-bonding ones (MOs 30 and 31) as well-as non-bonding $\pi$-orbitals (MOs 32 -- 35). The EFG given by the Dailey--Townes model is the \ce{Cl} principal contribution $\braket{e\hat{q}^{\text{el}}_{\text{Cl}}}_{\text{princ}}^{(\text{Cl})}$ 5.891 E$_{h}/ea^{2}_{0}$ arising from the valence orbitals. This number is slightly different from the value 5.902 E$_{h}/ea^{2}_{0}$ reported Table \ref{tab:we}; this is because the net populations reported in that Table are accumulated over all MOs of the molecule. It is of interest to note that the valence non-bonding $\sigma$-orbitals contribute -0.362 E$_{h}/ea^{2}_{0}$ to $\braket{e\hat{q}^{\text{el}}_{\text{Cl}}}_{\text{princ}}^{(\text{Cl})}$, although Table \ref{tab:ICl_local} suggests that MOs 30 and 31 essentially correspond to \ce{I}5s and \ce{Cl}3s, respectively. However, Table \ref{tab:ICl_local} also shows participation of \ce{Cl}3p$_z$ to these orbitals; it is minute, but enough to explain the cited contribution. 

Let us now consider what further contributions beyond the Dailey--Townes model that are needed to arrive at the total EFG of 4.736 $\text{E}_{h}/ea_{0}^{2}$. From Table \ref{tab:ICl_decomp} we see that the iodine intra-atomic contribution $\braket{e\hat{q}^{\text{el}}_{\text{Cl}}}_{\text{intra}}^{(\text{I})}$ from core and valence sum up to -1.202 $\text{E}_{h}/ea_{0}^{2}$, which is clearly not a small number. However, this contribution is to a large extent cancelled by the nuclear contribution of 1.255 $\text{E}_{h}/ea_{0}^{2}$. This can be understood from a Taylor-expansion of the operator $e\hat{q}_{K}^{e}$ about another center ($A\ne K$)
\begin{equation}
    \langle\psi_{p}^{A}|e\hat{q}_{K}^{e}|\psi_{q}^{A}\rangle=
    -\frac{e}{4\pi\varepsilon_{0}}\left[\frac{3Z_{AK}^{2}-R_{AK}^{2}}{R_{AK}^{5}}\right]\delta_{pq}+\sum_{\alpha}\left.\frac{\partial e\hat{q}_{K}^{e}}{\partial r_{A;\alpha}}\right|_{\mathbf{r}_{A;\alpha}=\mathbf{0}}\langle\psi_{p}^{A}|r_{A;\alpha}|\psi_{q}^{A}\rangle+\ldots
\end{equation}
The zeroth-order terms corresponds exactly to a nuclear contribution of Eq.\eqref{eq:EFGexp}, but with opposite sign. Further terms in the expansion involve matrix elements $\langle\psi_{p}^{A}|x_A^iy_A^jz_A^k|\psi_{q}^{A}\rangle$ of increasing Cartesian powers $i+j+k=\ell$ balanced against increasing inverse powers of $R_{AK}^{-(3+\ell)}$. In less mathematical terms, the electron cloud of the iodine atom, seen from the chlorine atom, appears as a point-like charge which thereby gives an electronic contribution of the same magnitude as the nuclear one, but of opposite sign.

Further inspection of Table \ref{tab:ICl_decomp} shows that there is a \ce{Cl} hybridization contribution $\braket{e\hat{q}^{\text{el}}_{\text{Cl}}}_{\text{hyb}}^{(\text{Cl})}$ from valence of -0.616 $\text{E}_{h}/ea_{0}^{2}$, most of which comes from the bonding $\sigma$-orbital. This is somewhat surprising, since Eq.~\eqref{eq:orb29} is a faithful description of this orbital. However, our projection analysis shows that there is a  contribution from \ce{Cl}2p$_z$, with a very small coefficient of $-5.94\cdot 10^{-2}$, but it multiplies a large matrix element (
$\langle\text{Cl}3p_z|e\hat{q}_{K}^{e}|\text{Cl}2p_z\rangle=21.182 \text{E}_{h}/ea_{0}^{2}$) and the \ce{Cl}3p$_z$ coefficient ($a=0.703$) to give an important contribution.

A final significant contribution is that of polarization $\braket{e\hat{q}^{\text{el}}_{\text{Cl}}}_{\text{pol}}^{(\text{Cl})}$, summing up to -0.619 E$_{h}/ea^{2}_{0}$. When projection analysis is used to decompose the integrated number density according to Eq.~\eqref{eq:popana}, significant polarization is an indication that some important atomic orbital contribution is missing. However, this is not the case for the present systems. The polarization contribution is only large for the electronic EFG expectation value and not for populations. We will therefore claim that the polarization contribution provides a precise definition of the deformation of atomic densities that occur when atoms are brought together to form a molecule.

\begin{table}[h]
  \centering
  \begin{tabular}{lccccclcc}
    \toprule
    & $\braket{e\hat{q}_{\ce{Cl}}^{\text{el}}}_{\text{intra}}^{(\ce{Cl})}$ & $\braket{e\hat{q}_{\ce{Cl}}^{\text{el}}}_{\text{princ}}^{(\ce{Cl})}$ & $\braket{e\hat{q}_{\ce{Cl}}^{\text{el}}}_{\text{hyb}}^{(\ce{Cl})}$ &$\braket{e\hat{q}_{\ce{Cl}}^{\text{el}}}_{\text{inter}}^{(\ce{Cl}X)}$ & $\braket{e\hat{q}_{\ce{Cl}}^{\text{el}}}_{\text{pol}}$  &$\braket{e\hat{q}_{\ce{Cl}}^{\text{el}}}^{(\text{X})}_{\text{intra}}$ & $e\hat{q}_{\ce{Cl}}^{\text{nucl}}$ & $\braket{e\hat{q}_{\ce{Cl}}}$ \\
    \midrule
    \ce{Cl2}   & 6.662 & 7.320  & -0.659  & -0.020 & -0.618  &-0.539 & 0.591 & 6.075 \\
    \ce{ICl}   & 5.319 & 5.890  & -0.571  & -0.017 & -0.619  &-1.202 & 1.255& 4.736\\
    \ce{ClCN}  & 5.114 & 6.223 & -1.109  & -0.062 & -0.852  &-0.364 & 0.504 & 4.340 \\
    \ce{CH3Cl} & 5.146 & 6.082 & -0.936  & -0.033 & -1.153  &-0.301 & 0.359 & 4.018 \\
    \ce{NaCl}  & 0.769 & 0.874  & -0.105  & 0.002 & -0.213  &-0.219 & 0.240 & 0.579 \\
    \bottomrule
  \end{tabular}
  \caption{Decomposition of the electric field gradient at the nuclear position of \ce{Cl} in molecules \ce{Cl\bond{-}X} using projection analysis. All values are expressed in atomic units (E$_{h}/ea_{0}^{2}$). $\braket{e\hat{q}_{\ce{Cl}}^{\text{el}}}^{(\text{X})}_{\text{intra}}$ is the sum of all intra-atomic contributions coming from the other atoms present in the molecule.}
  \label{tab:ClX_decomp}
\end{table}
%\begin{table}[H]
%    \centering
%    \begin{tabular}{l|l l |c|c}
%         & & &no-SO  &SO\\
%         \hline
%         intra-atomic& Cl& Principal & 5.890&1.741\\
%         & & Hybridization &-0.570 &3.496\\
%         \cline{2-5}
%         &I&  Principal    &-1.212&-1.146\\
%         &       & Hybridization & 0.010&0.019\\
%         \hline
%         inter-atomic & Cl-I & & -0.017&-0.016\\
%         \hline
%         Polarization & & &-0.619&-0.575\\
%         \hline
%         Total electronic &  & &3.482&3.518\\
%         \hline
%         Total nuclear &  & &1.254&1.174\\
%         \hline
%         Total EFG &  & &4.737 &4.693\\
%         \hline
%    \end{tabular}
%    \caption{Decomposition of EFG at the position of \ce{^{35}Cl} in \ce{ICl} with and without %spin-orbit interaction (SO). All values are reported in $\text{E}_{h}a_{0}^{-2}$}
% \label{tab:tabICl}
%\end{table}
Our observations concerning the \ce{ICl} molecule are valid for all the studied molecules, as seen from Table \ref{tab:ClX_decomp}. We note that:
\begin{itemize}
\item The expectation value $\braket{e\hat{q}_{\ce{Cl}}}$ is dominated by the intra-atomic contribution $\braket{e\hat{q}_{\ce{Cl}}^{\text{el}}}_{\text{intra}}^{(\ce{Cl})}$, in particular the principal contribution $\braket{e\hat{q}_{\ce{Cl}}^{\text{el}}}_{\text{princ}}^{(\ce{Cl})}$, which is essentially the Dailey--Townes model.
\item The hybridization contribution $\braket{e\hat{q}_{\ce{Cl}}^{\text{el}}}_{\text{hyb}}^{(\ce{Cl})}$ can be significant. It arises from mixing of valence $p$ orbitals with subvalence ones, even by very small amounts, and typically reduces the Dailey--Townes value by 10-15 \%.
\item Inter-atomic contributions $\braket{e\hat{q}_{\ce{Cl}}^{\text{el}}}_{\text{inter}}^{(\ce{Cl}X)}$ are generally small and can be ignored. They therefore do not distinguish between the Dailey--Townes model\cite{Dailey_JCP1949} and the refinements proposed by Cotton and Harris.\cite{Cotton_Harris_PNAS1966}
\item Intra-atomic contributions  $\braket{e\hat{q}_{\ce{Cl}}^{\text{el}}}^{(\text{X})}_{\text{intra}}$ from other centers are not negligible, but to a large extent cancelled by corresponding nuclear contributions $e\hat{q}_{\ce{Cl}}^{\text{nucl}}$. We may expect that the combined contribution can be calculated as a nuclear contribution, but replacing nuclear charges by atomic partial ones.
\item Polarization contributions $\braket{e\hat{q}_{\ce{Cl}}^{\text{el}}}_{\text{pol}}$ may be important and have a clear physical interpretation.
\end{itemize}

So far we only considered scalar relativistic effects in our calculations. Let us now investigate the effect of spin-orbit coupling. For $p$-shells the non-zero angular factors of Eq.~\eqref{eq:angfac} are $A_{+-;\pm 1/2}=\mp\sqrt{2}/5$,  $A_{++,\pm 1/2}=1/5$ and $A_{++,\pm 3/2}=-1/5$. In Table \ref{tab:Cl_SO} we show atomic matrix elements $\braket{e\hat{q}_{Cl}}_{ii}$ obtained in the same manner as in Table \ref{tab:tabpol}, but now with spin-orbit interaction included. The values are consistent with the expression given in Eq.~\eqref{eq:matrix_SO}. For instance, the matrix elements for 3p$_{3/2;1/2}$ and 3p$_{3/2;3/2}$ both imply an inverse cubic radial expectation value $\langle r^{-3}\rangle_{3p;++}=7.455 a_0^{-3}$. As already discussed, the expectation value $\langle r^{-3}\rangle_{3p;--}$ is not accessible from the EFG due to spherical symmetry of the p$_{1/2}$ orbital density. Direct calculation gives $\langle r^{-3}\rangle_{3p;--}=10.850 a_0^{-3}$, which is a 45\% increase with respect to $\langle r^{-3}\rangle_{3p;++}$. In comparison the radial expectation value $\langle r\rangle_{3p;--}=1.834 a_0$ is only -0.53\% smaller than $\langle r\rangle_{3p;+-}=1.844 a_0$. The EFG operator, due to its inverse cubic radial dependency, is clearly very sensitive to changes in orbital size due to spin-orbit interaction. This is the origin of the spin-orbit tilting discussed by Pyykk{\"o} and Seth.\cite{Pyykkoe_TCA1997} 
\begin{table}[H]
    \centering
    \begin{tabular}{ l c}
         Orbital & $\braket{e\hat{q}_{Cl}}_{ii}$ \\
         \hline
         1s$_{1/2;1/2}$&  0.000\\ 
         \hline
         2s$_{1/2;1/2}$&   0.000\\
         2p$_{1/2;1/2}$&  0.000\\
         2p$_{3/2;1/2}$&   -40.500\\
         2p$_{3/2;3/2}$&   40.500\\
         \hline
         3s$_{1/2;1/2}$& 0.000\\
         3p$_{1/2;1/2}$& 0.000\\
         3p$_{3/2;1/2}$& -2.982 \\
         3p$_{3/2;3/2}$&  2.982 \\
    \end{tabular}
    \caption{\footnotesize Atomic matrix elements $\braket{e\hat{q}_{Cl}}^{Cl}$ in the presence of spin-orbit coupling. All values of $eq_{Cl}$ are in atomic units (E$_{h}/ea_{0}^{2}$). }
    \label{tab:Cl_SO}
\end{table}

\begin{table}[H]
  \centering
  \begin{tabular}{lccccclcc}
    \toprule
    & $\braket{e\hat{q}_{\ce{Cl}}^{\text{el}}}_{\text{intra}}^{(\ce{Cl})}$ & $\braket{e\hat{q}_{\ce{Cl}}^{\text{el}}}_{\text{princ}}^{(\ce{Cl})}$ & $\braket{e\hat{q}_{\ce{Cl}}^{\text{el}}}_{\text{hyb}}^{(\ce{Cl})}$ &$\braket{e\hat{q}_{\ce{Cl}}^{\text{el}}}_{\text{inter}}^{(\ce{Cl}X)}$ & $\braket{e\hat{q}_{\ce{Cl}}^{\text{el}}}_{\text{pol}}$  &$\braket{e\hat{q}_{\ce{Cl}}^{\text{el}}}^{(\text{X})}_{\text{intra}}$ & $e\hat{q}_{\ce{Cl}}^{\text{nucl}}$ & $\braket{e\hat{q}_{\ce{Cl}}}$ \\
    \midrule
    \ce{ICl}   & 5.319 & 5.890  & -0.571  & -0.017 & -0.619  &-1.202 & 1.254 & 4.735 \\
               & 5.237 & 1.741  &  3.496  & -0.016 & -0.575  &-1.127 & 1.174 & 4.693 \\
    \ce{AtCl}& 5.073& 5.595& -0.522& -0.016& -0.608&-1.733& 1.780& 4.495\\
    & 4.365& 1.233& 3.132& -0.010& -0.544& -1.534& 1.567& 3.844\\
    \ce{TsCl}& 4.690& 5.177& -0.487& -0.015& -0.615&-2.122& 2.165& 4.102\\
 & 3.050& 0.773& 2.276& -0.007& -0.506& -1.713& 1.739&2.561\\
  \end{tabular}
  \caption{Decomposition of the electric field gradient at the nuclear position of \ce{Cl} in molecules \ce{Cl\bond{-}X} using projection analysis. All values are in atomic units (E$_{h}/ea_{0}^{2}$). The second line for each molecule gives the decomposition upon inclusion of spin-orbit interaction.}
  \label{tab:Cl_ClX_decomp}
\end{table}
The mixed value $\langle r^{-3}\rangle_{3p;+-}=7.542 a_0^{-3}$ does not contribute to atomic expectation values, but can be expected to contribute significantly to molecular expectation values and, in our terminology, as a hybridization contribution. For instance, using the notation of Pyykk{\"o} and Seth,\cite{Pyykkoe_TCA1997} the contribution from chlorine to the $\sigma$-bond of \ce{ICl} can be expressed as
\begin{equation}
    3p_{\sigma} (1/2) = -\sqrt{\frac{1}{3}}3p_{1/2;1/2}+\sqrt{\frac{2}{3}}3p_{3/2;1/2},
\end{equation}
giving rise to an EFG expectation value
\begin{equation}
    \langle 3p_{\sigma} (1/2)|e\hat{q}_{\ce{Cl}}^{e}|3p_{\sigma} (1/2)\rangle = - \frac{2e}{4\pi\varepsilon_0}\left(\frac{4}{15}\langle r^{-3}\rangle_{3p;+-}+\frac{2}{15}\langle r^{-3}\rangle_{3p;++}\right).
\end{equation}
Note, however, that this reduces to the scalar-relativistic expectation value
\begin{equation}
    \langle 3p_z \alpha|e\hat{q}_{\ce{Cl}}^{e}|3p_z \alpha\rangle = - \frac{2e}{4\pi\varepsilon_0}\times\frac{2}{5}\langle r^{-3}\rangle_{3p}
\end{equation}
only when the radial matrix elements $\langle r^{-3}\rangle_{p;+-}$ and $\langle r^{-3}\rangle_{p;++}$ coincide. 

In Table \ref{tab:Cl_ClX_decomp} we show the decomposition of the calculated electric field gradient at the nuclear position of \ce{Cl} in the interhalogens \ce{X\bond{-}Cl}, (X=\ce{I}, \ce{At}, \ce{Ts}), without and with spin-orbit interaction included. We recall that the calculations were carried out at the corresponding optimized equilibrium distances, as seen in Table \ref{tab:bondX}. When spin-orbit interaction is included, the hybridization contribution $\braket{e\hat{q}_{\ce{Cl}}^{\text{el}}}_{\text{hyb}}^{(\ce{Cl})}$ is systematically larger than the principal contribution $\braket{e\hat{q}_{\ce{Cl}}^{\text{el}}}_{\text{princ}}^{(\ce{Cl})}$. Closer inspection shows that both contributions are dominated by the $3p$-shell of chlorine. In \ce{ICl} the sum of the two contributions, the total intra-atomic contribution $\braket{e\hat{q}_{\ce{Cl}}^{\text{el}}}_{\text{intra}}^{(\ce{Cl})}$, is little affected by spin-orbit interaction and the redistribution between principal and hybdridization contributions simply reflects the change of coupling regime of the underlying atomic reference orbitals. However, replacing iodine by the heavier halogens astatine and tennessine a genuine spin-orbit effect on $\braket{e\hat{q}_{\ce{Cl}}^{\text{el}}}_{\text{intra}}^{(\ce{Cl})}$ comes into play, induced by the heavier atom, as also seen on the optimized bond distances in Table \ref{tab:bondX}. Furthermore, as seen in Table \ref{tab:X_ClX_decomp}, if we instead focus on the electric field gradient at the heavier atom \ce{X} of these compounds, the hybridization contribution is overtaken by the principal contribution. Finer scrutiny shows that these contributions are dominated by the valence $p$-shell of the heavy atom. Keeping in mind that the EFG operator probes the vicinity of the target atom, this can be seen as the transition from LS coupling to jj coupling on the heavy atom. It should also be noted that the decomposition is invariant to a rotation among the occupied molecular orbitals. We may also note that a bond length extension of 7.5\% for \ce{TsCl}, induced by spin-orbit interaction, translates into a 19.7\% reduction of the nuclear EFG-contributions $e\hat{q}_{\ce{Cl}}^{\text{nucl}}$ and $e\hat{q}_{\ce{X}}^{\text{nucl}}$, again showing the sensitivity of the EFG-operator.

\begin{table}[H]
  \centering
  \begin{tabular}{lccccclcc}
    \toprule
    & $\braket{e\hat{q}_{\ce{X}}^{\text{el}}}_{\text{intra}}^{(\ce{X})}$& $\braket{e\hat{q}_{\ce{X}}^{\text{el}}}_{\text{princ}}^{(\ce{X})}$& $\braket{e\hat{q}_{\ce{X}}^{\text{el}}}_{\text{hyb}}^{(\ce{X})}$&$\braket{e\hat{q}_{\ce{X}}^{\text{el}}}_{\text{inter}}^{(\ce{Cl}X)}$& $\braket{e\hat{q}_{\ce{X}}^{\text{el}}}_{\text{pol}}$&$\braket{e\hat{q}_{\ce{X}}^{\text{el}}}^{(\text{Cl})}_{\text{intra}}$& $e\hat{q}_{\ce{X}}^{\text{nucl}}$& $\braket{e\hat{q}_{\ce{X}}}$\\
    \midrule
    \ce{ICl}   & 19.772& 22.052& -2.281& -0.031& -1.234&-0.382& 0.402& 18.527\\
               & 19.012& 8.317&  10.694& -0.027& -0.960&-0.360& 0.376& 18.041\\
    \ce{AtCl}& 40.908& 45.453& -4.544& 0.011& -1.957&-0.342& 0.356& 38.976\\
    & 31.424& 17.386& 14.038& 0.013& -1.002& -0.306& 0.314& 30.443\\
    \ce{TsCl}& 91.377& 101.123& -9.744& -0.039& -3.832&-0.305& 0.314& 87.515\\
 & 44.895& 28.910& 15.984& -0.025& -0.781& -0.252& 0.252&44.089\\
  \end{tabular}
  \caption{Decomposition of the electric field gradient at the nuclear position of \ce{X} in molecules \ce{X\bond{-}Cl} using projection analysis. Each value is expressed in atomic units (E$_{h}/ea_{0}^{2}$). The second line for each molecule gives the decomposition upon inclusion of spin-orbit interaction.}
  \label{tab:X_ClX_decomp}
\end{table}

Finally, before closing this Section, we summarize in Table \ref{tab:soc} calculated values of the EFG at the \ce{Cl} position in different dihalogens using different Hamiltonians. Starting from non-relativistic values, the inclusion of scalar-relativistic effects increase the EFG, whereas the further inclusion of spin-orbit effects reduces it. For the heaviest systems the overall relativistic effect is a reduction of the non-relativistic value. These results are in line with those obtained by Aucar \textit{et al.} at both the HF and DFT level.\cite{Aucar_IJQC2021}

\begin{table}
\begin{tabular}{l|r|r|r|r|r|r}
&$eq^{NR}_{\ce{Cl}}$ & $eq^{SR}_{\ce{Cl}}$ & $eq^{SO}_{\ce{Cl}}$ & $\Delta_{SR-NR}$ & $\Delta_{SO-SR}$ & $\Delta_{SO-NR}$ \\
\hline
\ce{FCl}&7.622&7.709&7.707&0.087&-0.002&0.085\\
\ce{Cl2}&6.007&6.076&6.037&0.069&-0.039&0.03\\
\ce{BrCl}&5.502&5.567& 5.532&0.065&-0.035&0.03\\
\ce{ICl}&4.649&4.736&4.693&0.088&-0.044&0.044\\
\ce{AtCl}&4.318&4.496&3.844&0.178&-0.652&-0.474\\
\ce{TsCl} & 3.719 & 4.102& 2.561&0.383&-1.541&-1.158\\
\end{tabular}
\caption{Calculated values in E$_{h}/ea^{2}_{0}$ of the electric field gradient at the \ce{Cl} position in different dihalogens. The values are reported at the non-relativistic level ($eq^{NR}_{\ce{Cl}}$), adding scalar relativistic effects ($eq^{SR}_{\ce{Cl}}$) and finally also  spin-orbit interaction ($eq^{SO}_{\ce{Cl}}$).}
\label{tab:soc}
\end{table}

\subsection{Uranyl}
In order to go beyond the realm of the Dailey--Townes model, we have chosen to study the electric field gradient at the position of uranium in uranyl [\ce{UO2^2+}]. In this widely studied molecule\cite{pepper:actinides,Denning_ch1992,Denning_JPCA2007,Kovacs_CR2015} bonding is mediated primarily by the uranium 5$f$ and 6$d$ orbitals. An intriguing feature of the electronic structure of uranyl is the presence of the so-called 
6$p$-hole. Veal and co-workers in 1975 reported an X-ray photoemission spectroscopy study of hexavalent uranium compounds, most of which contained the uranyl unit, and observed a splitting of the 6$p_{3/2}$ level that depended strongly on the separation between uranium and the nearest-neighbour oxygens, and which they therefore attributed to ligand-field splitting.\cite{Veal_PRB1975} Subsequent calculations showed that the interaction is even stronger;\cite{Walch_Ellis_JCP1976,Yang_JCP1978,DeKock_CPL1984,Tatsumi_IC1980,Pyykko_IC1981,Pyykko_JPC1984,Pyykko_CP1986,Pyykko_ICA1987} \ce{U} $6p$ and \ce{O} 2s combine to form molecular orbitals. This is also 
in line with Grechukhin and co-workers who attributed the presence of a \ce{O} $2s$ peak in the \ce{^235U} internal conversion spectrum of uranium trioxide to appreciable hybridization of these orbitals with \ce{U} $6p$.\cite{Grechukin_PZETZ1980} Further insight was provided by a 
polarized X-ray spectroscopic study of a single crystal of \ce{Cs2UO2Cl4} by Denning and co-workers, where a weak band in the \ce{O} K$\alpha$ emission spectrum was interpreted as revealing O $2p$ character in the $\sigma_{1/2}$ component of the \ce{U} $6p_{3/2}$ orbital.\cite{Denning_JCP2002,Denning_JPCA2007}

At the relativistic Hartree--Fock level, the radial expectation value of uranium $6p$ extends beyond that of $5f$.\cite{Desclaux_ADNDT1973} The subvalence character of the \ce{U} $6p$ orbital has been invoked to explain the relativistic bond length \textit{extension} of uranyl,\cite{vanWezenbeek_TCA1991} as well as the observation that the molecule is linear,\cite{Tatsumi_IC1980} in contrast to the bent structure of transition metal dioxo ions, as well as isoelectronic \ce{ThO2}, although the latter explanation has been refuted by Wadt.\cite{Wadt_JACS1981} Of particular interest in the present work is the 1986 suggestion by Larsson and Pyykkö that the electric field gradient at the position of uranium in uranyl arises essentially from the 6$p$-hole. 

The NQCC has not been measured for the bare uranyl unit, but has been extracted from the Mössbauer spectrum of the $2^+\rightarrow0^+$ $\gamma$-transition of \ce{^{234}U} in nonmagnetic uranyl rubidium nitrate \ce{[(UO2)Rb(NO3)3]}; assuming axial symmetry. Monard \textit{et al.}  report $(3/8)e^2qQ=(-60.1\pm 0.3)$ mm/s.\cite{Monard_PhysRevB.9.2838} The authors also provided an estimate of $eq$: The intrinsic nuclear quadrupole moment $Q_0$ can be extracted from the electric quadrupole transition probability $B(E2;0^+\rightarrow2^+)$. For \ce{^{234}U}, Ford \textit{et al.} reported $Q_0=(10.19\pm 0.13)$ b.\cite{Ford_PhysRevLett.27.1232} Assuming an axially symmetric nuclear deformation with the nuclear spin having a well-defined direction with respect to the symmetry axis of the deformation (strong coupling\cite{Neugart_Neyens_2006,Moszkowski_HP1957}), this translates into a spectroscopic (molecular-frame) quadrupole moment of $Q=(-2.91\pm 0.04)$ b. Using this value, Monard \textit{et al.} suggested an electric field gradient $eq=(8.14\pm 0.13)\times 10^{18}$ V/cm$^2$ (corresponding to $(8.38\pm 0.13)\ E_h/ea_0^2$). Repeating their calculation, we arrive at a somewhat lower electric field gradient, $eq=7.99\times10^{18}$ V/cm$^2$ (or 8.22 $E_h/ea_0^2$). Reproducing their value requires selecting a $\gamma$-transition energy $E_{\gamma}$=4.431 keV instead of the value $E_{\gamma}$=4.3491 keV used by the authors and taken from Ref. \citenum{Schmorak_NPA1972} (the modern value is $E_{\gamma}$=4.34981 keV\cite{Pritychenko_ADNDT2016}). 

Larsson and Pyykkö made the following suggestion:\cite{Pyykko_CP1986} Suppose that the entire electric field gradient of uranyl rubidium nitrate [\ce{(UO2)Rb(NO3)3}] comes from the diagonal element $\langle e\hat{q}^e_U\rangle_{ii}$ with $i=U\ 6p_{3/2;1/2}$, that they in atomic units express as $(3/5)\langle r^{-3}\rangle_{++}$, and that notably Sternheimer shielding (core polarization) is ignored. Taking the HF radial expectation value\cite{Desclaux_ADNDT1973} $\langle r^{-3}\rangle_{++}=80.61a_0^{-3}$ , they then found that the matrix element evaluates to 48.37 $E_h/ea_0^2$ and that this would require a $6p_{3/2;1/2}$-hole of $0.17e$. In a subsequent paper Pyykkö and Seth point out that the angular factor was wrong;\cite{Pyykkoe_TCA1997} it should be $(2/5)\langle r^{-3}\rangle_{++}$, now requiring a $6p_{3/2;1/2}$-hole of $0.26e$. Strictly speaking, though, the matrix element comes with a negative sign; it should be
\begin{equation}
   \langle U\ 6p_{3/2;1/2}|e\hat{q}^e_U|U\ 6p_{3/2;1/2}\rangle=-\frac{2e}{4\pi\varepsilon_0} \langle r^{-3}\rangle_{++}\times (1/5)
\end{equation}
This disagrees with the sign of $eq$ as reported by Monard and co-workers. There are many instances, with \ce{^{57}Fe} as a prominent example, where the Mössbauer spectrum is invariant under a change of sign of $eq$, and so the sign of $eq$ can only be deduced by special techniques, such as the application of a magnetic field.\cite{gutlich_Mossbauer} However, in the case of \ce{^{234}U} the $2^+\rightarrow0^+$ $\gamma$-transition gives a three-line Mössbauer spectrum with an energy splitting ratio of 3:1 and relative line intensities of 2:2:1,\cite{Monard_PhysRevB.9.2838,Shenoy_NIM1969}and so there is indeed a problem of sign. A 4-component relativistic HF study by de Jong and co-workers\cite{deJong_THEOCHEM1998} showed a positive value ($eq=2.7\ E_h/ea_0^2$), with important contributions beyond that of the $U(6p)$ core-hole, although the latter dominated the distance-dependence of the EFG, in line with the suggestion by Larsson and Pyykkö.\cite{Pyykko_CP1986}  Upon inclusion of negative point charges to simulate the effect of equatorial ligands, the authors observed a significant increase ($eq=6.6\ E_h/ea_0^2$). These calculations employed a default \ce{U\bond{-}O} bond length of 1.78 Å, in line with the experimental one for \ce{[(UO2)Rb(NO3)3]}. Subsequent calculations have shown that the inclusion of electron correlation by the Kohn–Sham approach invariably makes the electric field gradient \textit{negative}, whereas the inclusion of explicit equatorial ligands brings it back to a positive value.\cite{Belanzoni2005,Aquino_JCTC2010,Autschbach_JCTC2012} The latter observation was also confirmed by a recent CCSD/X2C-AMFI study on the uranyl tris-nitrate complex (\ce{[(UO2)(NO3)3]-}), albeit using a rather small basis.\cite{pototschnig_jctc2021}

Our investigation begins with an assessment of the electronic structure of uranyl. In Table \ref{tab:conf} the electronic configurations of the atoms within the molecule in a purely ionic perspective is contrasted with what is actually extracted from our PBE calculations using projection analysis. In the latter case, we see that the effective charge on uranium is +2.46e, with and without SO, whereas the oxygens each carry small negative charges of -0.08e. There is, however, some uncertainty in these partial charges since the polarization contribution is -0.30e and represents charge not attributed to any center. Switching to intrinsic atomic orbitals (IAOs), \cite{Knizia_JCTC2013} the polarization contribution is eliminated, and the partial charges are +2.38e and -0.19e for U and O, respectively. Looking at the effective electron configuration of uranium in the molecule, we find that the U(5$f$) and U(6$d$) orbitals exhibit partial occupation and, most notably, we observe a U(6$p$)-hole of 0.28e; upon inclusion of spin-orbit interaction, most of the hole (0.25e) is located on the U(6$p_{3/2}$) orbital.
\begin{table}[h]
    \centering
    \begin{tabular}{l|l c c l}
       Atom & &Charge  &  &Occupation of atomic orbitals\\
       \hline
       \hline
       U & Ionic& +6  &  & [Hg]6$p^{6}$5$f^{0}$6$d^{0}$7$s^{0}$  \\
         & Calculated& +2.46 &  & [Hg]6$p^{5.72}$5$f^{2.74}$6$d^{1.05}$7$s^{0.05}$\\
         & -- with SOC&+2.46 &  &[Hg]6$p_{1/2}^{1.94}$6$p_{3/2}^{3.75}$5$f_{5/2}^{1.23}$5$f_{7/2}^{1.49}$6$d_{3/2}^{0.48}$6$d_{5/2}^{0.60}$7$s_{1/2}^{0.05}$\\
       \hline
       O & Ionic& -2 &  &  1$s^{2}$2$s^{2}$2$p^{6}$ \\
         & Calculated& -0.08 &  & 1$s^{2}$2$s^{1.9}$2$p^{4.41}$\\
    \end{tabular}
    \caption{Analysis of the electronic structure of \ce{UO2^2+} from the purely ionic point of view to that obtained from the results conducted with DFT using projection analysis. The polarization contribution is -0.30e.}
    \label{tab:conf}
\end{table}

Table \ref{tab:canonMOs} shows the distribution of the U(6$p$)-hole over molecular orbitals, again using projection analysis. We do indeed observe significant mixing of U(6$p$) with O(2$s$/2$p$)-orbitals in MOs $\sigma_{2},\sigma_{4}$ and $\sigma_{7}$, which, for instance, is in line with the observations of Denning and co-workers.\cite{Denning_JCP2002,Denning_JPCA2007} However, these are \textit{canonical} MOs, with well-defined energies and symmetries and, as stressed by Mulliken,\cite{PhysRev.41.49} although such spectroscopic MOs, as he called them, are well suited for the understanding of electronic spectra and ionization, they do not allow one to "see" chemical bonds. In order to describe and understand chemical bonding, Mulliken advocated the use of maximally localized MOs, that is, in his terminology, \textit{chemical} MOs.  
\begin{table}[H]
    \centering
    \begin{tabular}{l c | c c c c c|c c|c c}
         & $\varepsilon$ (E$_{h}$) & 6s &6p &5f &6d &7s & \ce{O1} 2s& \ce{O1} 2p&  \ce{O2} 2s& \ce{O2} 2p  \\
         \hline
         \hline
         $\sigma_{1}$& -2.343 &1.926& & & & &  0.019& 0.018& 0.019&   0.018\\
         $\sigma_{2}$&-1.644 & &1.080&0.035 & & & 0.349& 0.089&0.349& 0.089\\
         $\pi_{1}$&-1.432 &  & 1.964& & & & & 0.016& &   0.016\\
         $\pi_{2}$& -1.432&  & 1.964& & & & & 0.016& &   0.016\\
         $\sigma_{3}$&-1.374 &0.044& & &0.155& 0.002&0.850& 0.023&0.850&   0.023\\
         $\sigma_{4}$& -1.124&  &0.530&0.080& & & 0.593&0.098& 0.593&0.098\\
         $\pi_{3}$&-0.863 &  & & & 0.389& & & 0.767& &0.767\\
         $\pi_{4}$& -0.863&  & & & 0.389& & & 0.767& &0.767\\
         $\pi_{5}$&-0.851 &  & 0.022& 0.730& & & & 0.620& & 0.620\\
         $\pi_{6}$&-0.851 &  & 0.022& 0.730& & & & 0.620& & 0.620\\
         $\sigma_{6}$& -0.833 &0.017& & &0.084&0.044&0.077&0.821&0.077& 0.821\\
         $\sigma_{7}$&-0.806&  &  0.142& 1.162& & & 0.018&0.324& 0.018& 0.324\\
         \hline
         Sum &  & 1.987&5.724&2.737&1.017&0.046&1.906&4.179&1.906&  4.179\\
    \end{tabular}
    \caption{Gross population from projection analysis in canonical orbitals of uranyl.}
    \label{tab:canonMOs}
\end{table}
We therefore carried out localization using the Pipek-Mezey criterion,\cite{Pipek-Mezey_JCP1989}  which minimizes the number of centers spanned by  each MO. Somewhat to our surprise, we found that the localization procedure also induced rotations involving core orbitals, although these are already perfectly localized. After investigation of the gradient of the Pipek-Mezey functional with respect to orbital rotation, as detailed in Appendix \ref{app:loc}, we found that this is actually not excluded. As a pragmatic solution we therefore limited the localization procedure to the 12 canonical MOs of Table \ref{tab:canonMOs}. The gross populations, from projection analysis, of the resulting localized orbitals are shown in Table \ref{tab:locMOs}. The orbitals have been ordered according to approximate orbital energies $ \braket{\varepsilon}$, obtained as expectation values of the converged Kohn--Sham operator. Table \ref{tab:locMOs} suggests triple bonds between uranium and each oxygen (O$_1$:$\{\sigma_5,\pi_3,\pi_4\}$, O$_2$:$\{\sigma_6,\pi_5,\pi_6\}$, mainly involving U(5$f$)/U(6$d$) and O(2$p$), albeit with a non-negligible contribution from U(6$p$). This is in line with previous studies.\cite{Aquino_JCTC2010,bondUO2} Interestingly, the U(6$p$) contribution is significantly reduced upon inclusion of spin-orbit interaction, underlining the necessity of also considering the role of SO when investigating bonds involving heavy elements, as discussed by Knecht \textit{et al.} in the context of the \ce{U2} dimer.\cite{U2_dimer} 
\begin{table}[h]
    \centering
    \begin{tabular}{l c |c c c c c|c c|c c}
    \hline
        & $ \braket{\varepsilon}$& 6s &6p &5f &6d &7s & \ce{O1} 2s& \ce{O1} 2p&  \ce{O2} 2s& \ce{O2} 2p  \\
         \hline
         \hline
         $\sigma_{1}$& -3.429&1.021&0.791&0.251& 0.010& & -0.021& -0.054& &   \\
         $\sigma_{2}$& -3.429&1.021&0.791&0.251& 0.010& &  & & -0.021&    -0.054\\
         $\pi_{1}$&-2.827 &  & 1.990& 0.007& & & & -0.007& &   \\
         $\pi_{2}$&-2.827 &  & 1.990& 0.007& & & & & &   -0.007\\
         $\sigma_{3}$&-2.577 &  -0.003& -0.023& -0.003& 0.004& 0.018&1.921& 0.083& &   \\
         $\sigma_{4}$&-2.577 &  -0.003& -0.023&  -0.003& 0.004& 0.018& & & 1.921&   0.083\\
         $\sigma_{5}$&-2.116&  -0.024&0.108&0.391& 0.123& 0.004& 0.009& 1.322& &   0.021\\
         $\sigma_{6}$&-2.116 &  -0.024&0.108&0.391& 0.123& 0.004& & 0.021& 0.009&  1.322\\
         $\pi_{3}$&-1.733 &  &  -0.002&0.354& 0.195& & & 1.404& &  0.002\\
         $\pi_{4}$&-1.733&  &  -0.002&0.354& 0.195& & & 1.404& &  0.002\\
         $\pi_{5}$&-1.733&  &  -0.002&0.354& 0.195& & & 0.002& &  1.404\\
         $\pi_{6}$&-1.733 &  &  -0.002&0.354& 0.195& & & 0.002& &  1.404\\
         \hline
         Sum& & 1.988&5.724&2.708&1.054&0.044&1.909&4.177&1.909&  4.177\\
    \end{tabular}
    \caption{Gross population from projection analysis with localized orbitals in uranyl. $ \braket{\varepsilon}$ refers to the expectation value (in $E_{h}$)  of the converged Kohn-Sham operator.}
    \label{tab:locMOs}
\end{table}
\begin{table}[H]
    
    \centering
    \setlength{\tabcolsep}{2pt}
    \begin{tabular}{l l l|cccccccc|ccc|ccc}
        \toprule
        & $\omega$& $ \braket{\varepsilon}$& 6s$_{1/2}$ & 6p$_{1/2}$ & 6p$_{3/2}$ & 5f$_{5/2}$ & 5f$_{7/2}$ & 6d$_{3/2}$ & 6d$_{5/2}$ & 7s$_{1/2}$ & \ce{O1} 2s$_{1/2}$ & 2p$_{1/2}$ & 2p$_{3/2}$ & \ce{O2} 2s$_{1/2}$ & 2p$_{1/2}$ & 2p$_{3/2}$ \\
        \midrule
        \midrule
       42 & 1/2  & -3.422& 1.021&0.203 & 0.587& 0.106&0.145 & &  0.006& & -0.021& -0.021& -0.034&  -0.001& & \\
       43 & 1/2 &-3.422 & 1.02 & 0.203& 0.587&0.106& 0.145& &  0.006& & -0.001& & & -0.021& -0.021& -0.034\\
      44  & 1/2 &-3.156 &  & 1.514& 0.480 & 0.004& 0.012& & & & & -0.003& -0.003& & -0.003& -0.003\\
       45 & 3/2 &-2.634 &  & & 1.984& 0.010& 0.017& & & & & & & & & \\
      46  & 3/2 &-2.577 &  & -0.005& -0.018& & & 0.002& 0.003& 0.019& 1.920& 0.030& 0.053& & & \\
       47 & 3/2 &-2.577 &  & -0.005& -0.018& & & 0.002& 0.003& 0.019& & & & 1.920& 0.030& 0.053\\
      48 & 1/2 & -2.027&  & -0.018& 0.061& 0.348& &0.190& & & 0.002& &  0.013& 0.006& & 1.314\\
      49 & 1/2 &-2.027 &  & -0.018&0.061& 0.348& &0.190& & & 0.006& & 1.314& 0.002& & 0.013\\
      50 & 1/2  & -1.822&  & -0.006& 0.024& 0.027& 0.323& 0.036&0.133 & & &1.385& 0.018& & 0.006& \\
      51 & 1/2 &-1.822 &  & -0.006& 0.024& 0.027& 0.323& 0.036&0.133 & & & 0.006& & & 1.385& 0.018\\
      52 &  3/2 & -1.731&  & & & 0.131&0.226 & 0.047&0.143 & & & & 0.003& & &1.410\\
      53 & 3/2  &-1.731 &  & & & 0.131&0.226 & 0.047&0.143 & & & & 1.410& & & 0.003\\
      Sum&  & &  2.042& 1.862&3.772&1.238&1.417&0.550&0.570&0.038&1.906& 1.397& 2.774& 1.906& 1.397&2.774\\
        \bottomrule
    \end{tabular}
    \caption{Gross population from projection analysis in localized orbitals in uranyl with SOC. $ \braket{\varepsilon}$ refers to the expectation value (in $E_{h}$)  of the converged Kohn-Sham operator.}
    \label{tab:locMOs_SOC}
\end{table}
We now turn to the analysis of the electric field gradient at the nuclear position of uranium in \ce{UO2^2+}. Table \ref{tab:tabgg} shows a decomposition of the total EFG from projection analysis, with and without spin-orbit interaction. It can be seen that the intra-atomic contribution from uranium is quite small, some 10-15\% of the total value, arising in part from significant cancellation between the principal and hybridization contribution. Interestingly, these contributions change sign upon inclusion of spin-orbit interaction; we shall return to this later on. In line with what we have already seen for the chlorine compounds, the nuclear contribution is practically eliminated by the principal contribution of the oxygen atoms: $0.958-2*0.457=0.044 \text{E}_{h}/ea_{0}^{2}$. The inter-atomic contributions are small, albeit not negligible. What is really striking in Table \ref{tab:tabgg} is the polarization contribution, which completely dominates the total EFG. We have previously pointed out that a significant polarization contribution obscures projection analysis and may indicate that further atomic orbitals should be included in the analysis. However, in the present context we argue that the polarization contribution provides a precise measure of the contribution to the EFG arising from the deformation of atoms upon forming the molecule. This is vividly brought out by Table \ref{tab:MOs-EFG-uranyl} which shows the contribution to the EFG from groups of localized MOs, specifying for each group the fraction arising from polarization. It can be seen that almost half of the polarization contribution comes from the core orbitals, which were not included in the localization procedure.  From the inverse cubic radial dependency of EFG matrix elements it follows that even slight core polarization may give significant contributions to the total EFG. For instance, the lowest occupied MO (LOMO), which is essentially $U(1s)$, contributes 0.108 E$_{h}/ea_{0}^{2}$. Inspection shows that the $U(1s)$ orbital is polarized by $d$ orbitals (\textit{ungerade} orbitals like $p$ are not allowed due to inversion symmetry) and that the EFG contribution arises from the matrix elements coupling $s$ with $d$.

\begin{table}[H]
    \centering
    \begin{tabular}{l|l l |c|c}
         & & &without SO  &with SO\\
         \hline
         \hline
         intra-atomic& U& Principal & 2.093&-3.501\\
         & & Hybridization &-2.673&2.365\\
         \cline{2-5}
         &\ce{O1}&  Principal    &-0.457&-0.454\\
         &       & Hybridization & 0.021&0.018\\
         \cline{2-5}
         & \ce{O2}& Principal &-0.457&-0.454\\
         & & Hybridization & 0.021&0.018\\
         \hline
         inter-atomic & \ce{O1}-U & & 0.296&0.299\\
         & \ce{O2}-U & & 0.296&0.299\\
         & \ce{O1}-\ce{O2} & & -0.002& -0.002\\
         &Total & & 0.590&0.596\\
         \hline
         Polarization & & &-6.400&-6.813\\
         \hline
         Total electronic &  & &-7.262&-8.226\\
         \hline
         Total nuclear &  & &\multicolumn{2}{c}{0.958}\\
         \hline
         Total EFG &  & &-6.304 &-7.269\\
         \hline
    \end{tabular}
    \caption{Decomposition of EFG at the position of uranium in \ce{UO_2^{2+}} with and without spin-orbit interaction (SO). All values are reported in E$_{h}/ea_{0}^{2}$}
 \label{tab:tabgg}
\end{table}

In Table \ref{tab:tabgg} we provide a different decomposition of the EFG, now in terms of groups of Pipek--Mezey localized MOs. We have also added Table \ref{tab:MOs-intra_hyb-uranyl} which gives a further decomposition of uranium intra-atomic contributions to each group of MOs. For the moment, let us focus in the numbers obtained without inclusion of SO: It will be seen from Table \ref{tab:tabgg} that about half of the polarization contribution (and the EFG) comes from the core region. Clearly, the large core polarization masks the information about chemical bonding suggested by the Dailey--Townes model. Continuing our analysis, the non-bonding $O(2p)$ (lone pairs) contributes -1.246 E$_{h}/ea_{0}^{2}$, of which -0.418 E$_{h}/ea_{0}^{2}$ comes from polarization and -0.676 E$_{h}/ea_{0}^{2}$ from a minute contribution from $U(6p)$ and $U(5p)$ (cf. Table \ref{tab:MOs-intra_hyb-uranyl}); as we have seen, the intra-atomic electronic oxygen contributions are essentially cancelled by the nuclear ones.
Table \ref{tab:tabgg} furthermore shows that the non-bonding $U(6p)$ gives a very large \textit{positive} contribution, whereas the \ce{U\bond{3}O} bonds a somewhat larger \textit{negative} contribution. In fact, their sum basically complements the contribution from core polarization to give the full EFG. Upon inclusion of SO the overall picture is very little changed, but, in Tables 
\ref{tab:tabgg} and \ref{tab:MOs-intra_hyb-uranyl} one observes an approximate permutation of principal and hybridization contributions. This mainly arises from the fact, already mentioned, that $p_{1/2}$ orbitals do not contribute to atomic EFG expectation values, only through coupling with other AOs, principally $p_{3/2}$.

\begin{table}[H]
    \centering
    \begin{tabular}{|l|c|c|c|c|}
    \hline
         &\multicolumn{2}{c|}{without SO}&\multicolumn{2}{c|}{with SO}\\\cline{2-5}
         & $\braket{e\hat{q}_{U}^{e}}_{ii}$ &Polarization &$\braket{e\hat{q}_{U}^{e}}_{ii}$ &Polarization\\
         \cline{2-5}
         Core& -3.287& -3.107 (94.55\%)& -3.552& -3.112 (87.61\%)\\
 Non-bonding U(6p)& 29.761&4.802 (16.12 \%)& 28.354&4.190(14.78\%)\\
          Non-bonding O(2$p$)& -1.246&-0.418 (33.60 \%)& -1.287&-0.446 (34.88 \%)\\
          \ce{U\bond{3}O_1}& -16.245& -3.838(23.63\%)& -15.871& -3.722 (23.44 \%)\\
         \ce{U\bond{3}O_2}& -16.245&-3.838 (23.63\%)& -15.871&-3.722 (23.44 \%)\\
         \hline
         Electronic: $\braket{e\hat{q}_{U}^{e}}$& -7.262& -6.400 (88.14\%)& -8.227& -6.811 (82.79\%)\\
         \hline
          Nuclear: ${e\hat{q}_{U}^{n}}$&\multicolumn{4}{c|}{0.958}\\
          \hline
         Total $\braket{e\hat{q}_{U}}$& -6.304&& -7.269&\\
         \hline
    \end{tabular}
    \caption{All contributions of $eq_{U}$ (in $E_{h}/ea_{0}^{2}$) from the localized MOs of \ce{UO2^2+}. The core region starts from U(1$s$) and ends at U(5$d$). }
    \label{tab:MOs-EFG-uranyl}
\end{table}

\begin{table}[H]
    \centering
    \begin{tabular}{|l|c|c|c|c|}
    \hline
         &\multicolumn{2}{c|}{without SO}&\multicolumn{2}{c|}{with SO}\\\cline{2-5}
         & $\braket{e\hat{q}_{U}^{e}}_{\text{princ}}$&$\braket{e\hat{q}_{U}^{e}}_{\text{Hyb}}$&$\braket{e\hat{q}_{U}^{e}}_{\text{princ}}$&$\braket{e\hat{q}_{U}^{e}}_{\text{Hyb}}$\\
         \cline{2-5}
         Core& 0.475& -0.372& 0.219& -0.376\\
 Non-bonding U(6p)& 18.813&6.058& 5.536&18.542\\
          Non-bonding O(2$p$)& -2.859&2.183& -0.885& 0.196\\
          U$\equiv$\ce{O1}& -7.167& -5.271& -4.184&  -7.999\\
         U$\equiv$\ce{O2}& -7.167&-5.271& -4.184&-7.999\\
         \hline
 Total& 2.095& -2.673& -3.498&2.364\\
 \hline
    \end{tabular}
    \caption{Principal and hybridization contributions  of $eq_{U}$ (in $E_{h}/ea_{0}^{2}$) from the localized MOs of uranyl. The core region starts from U(1$s$) and ends at U(5$d$). }
    \label{tab:MOs-intra_hyb-uranyl}
\end{table}

The numbers of Table \ref{tab:tabgg} can to some extent be compared with an analysis of uranyl (and uranyl complexes) in terms of natural localized molecular orbitals (NLMOs)\cite{Reed_JCP1985b} provided by Aquino and co-workers.\cite{Aquino_JCTC2010} The decomposition of the EFG expectation value reported in Table 10 of their paper is based on scalar-relativistic calculations using the zeroth-order regular approximation (ZORA), the B3LYP functional and a bond length of 1.78 \AA. Their total value $eq$=-7.406 $E_{h}/ea_{0}^{2}$ is comparable to our scalar-relativistic value of -6.304 $E_{h}/ea_{0}^{2}$. At this level of theory, the NLMOs split into doubly occupied Lewis NLMOs and vacant non-Lewis ones. The expectation value is then expressed as Eq.~\eqref{eq:elexp} in terms of the former set. In order to avoid "large opposing nuclear and electronic" \cite{Autschbach2010} contributions, the authors directly include nuclear contributions into the electronic expectation value through the substitution
\begin{equation}
    e\hat{q}_{K}^{e}\rightarrow e\hat{q}_{K}^{e} + \frac{1}{N}eq_{K}^{n}.
\end{equation}
However, in our opinion, this even distribution of nuclear contributions over NLMOs fails to bring out the near perfect cancellation of electronic and nuclear contributions from the atoms surrounding the nucleus of interest, as  revealed by our projection analysis. The EFG decomposition of Aquino \textit{et al.}\cite{Aquino_JCTC2010} agrees with ours in terms of sign. Magnitudes are indeed smaller, but this can not be explained alone by the inclusion of nuclear contributions, since it is significantly smaller (0.958 $E_{h}/ea_{0}^{2}$) than the individual differences we observe.

\section{Conclusions}\label{sec:concl}
This work presents a detailed study of the Dailey--Townes model of electric field gradients (EFG) at nuclear positions in molecular system and the chemical information that can be extracted from their knowledge. The basic premise of the model is the sensitivity of the EFG to deviations of spherical symmetry, such as induced by chemical bonding. Dailey and Townes proposed that the EFG could be estimated from the occupations and atomic EFG matrix elements of the valence AOs on the center of interest. Dailey and Townes had $p$-bonding in mind when they formulated their model, but it can also accommodate the dominant bonding contributions from $d$- and $f$-orbitals, as seen in uranyl. The Dailey--Townes model can be seen as a particular approximation to the decomposition of the EFG expectation value by projection analysis.\cite{Dubillard_JCP2006,bast:pncana_PCCP2011} We have carried out such analysis for a number of halogen compounds, many of them included in the 1949 paper by Dailey and Townes, as well as uranyl. Overall, we find that the Dailey--Townes model in its pure form is too rough an approximation. Replacing net populations by gross populations improves the agreement with total SCF values, but we have found no valid theoretical justification for this. In more detail, we find that the intra-atomic contributions to the electronic EFG operator $e\hat{q}_{K}^{e}$  from centers other than the center $K$ of interest is significant, but to a large extent cancelled by corresponding nuclear contributions, in line with the arguments of Dailey and Townes.  Inter-atomic contributions are generally small, at least for the studied halogen compounds. However, hybridization contributions from the same center ($K$), which do not contribute to atomic expectation values, can be significant (-15\%). With increasing nuclear charge and number of electrons, core polarization reach the same order of magnitude as the EFG contributions associated with bonding, as dramatically seen in uranyl. A particular feature of uranyl is the $6p$-hole. Upon localization, we find the uranium $6p$-orbitals to be mostly non-bonding, in particular upon inclusion of spin-orbit interaction. Their large positive contribution combined with the somewhat larger negative contribution from the \ce{U\bond{3}O} bonds combine to give the other half of the total EFG.

We have also investigated relativistic effects. The inclusion of scalar relativistic effect does not fundamentally change the premises of the model. However, with the inclusion of spin-orbit interaction, the electronic EFG operator $e\hat{q}_{K}^{e}$ is no longer diagonal in the space of atomic orbitals of given orbital angular momentum $\ell$. We note in particular that $p_{1/2}$-orbitals, having a spherically symmetric density, has a zero EFG expectation value. The coupling between orbitals of $j=\ell\pm 1/2$, such as $p_{1/2}$ and $p_{3/2}$ becomes important, as manifested by important hybridization contributions. One could imagine rotating back from $|j,m_j\rangle$ eigenfunctions to $|\ell,m_{\ell},s,m_s\rangle$ ones, but this is only possible when the radial functions of AOs of $j=\ell\pm 1/2$ are identical, which will progressively no longer be the case as one descends the periodic table, and, in the language of Pyykkö and Seth,\cite{Pyykkoe_TCA1997} lead to spin-orbit tilting.

We believe that our detailed analysis of the electric field gradient at nuclear positions will be useful when setting up a computational protocol for accurate calculations of the EFG, a necessary ingredient for the direct determination of nuclear electric quadrupole moments from experiment. There are still a number of elements for which a reference $eq$ is not yet established.\cite{Stone_Int2024} In our group we are therefore developing tools for the generation of high-order relativistic coupled cluster code for such calculations.\cite{Brandejs_arxiv2024}

\begin{acknowledgments}
This project was funded by the European Research Council (ERC) under the European Union’s Horizon 2020 research and innovation programme (grant agreement ID:101019907). This project has been (partially) supported through the grant NanoX no ANR-17-EURE-0009 in the framework of the ”Programme des Investissements d’Avenir”. This work was performed using HPC resources from CALMIP (Calcul en Midi-Pyrenées; project P13154). TS would like to thank Pekka Pyykkö (Helsinki), Nick Stone (Oxford), Leonid Skripnikov (St. Petersburg) and Daniel SantaLucia (Mülheim) for helpful discussions. 
\end{acknowledgments}

\section*{Author declarations}
\subsection*{Conflict of Interest}
The authors have no conflicts to disclose.
\section*{Data Availability Statement}
The data that support the findings of this study are openly available in ZENODO at: ...
\newpage

%\section*{Support information}
%\begin{table}[h]
%\footnotesize
%    \centering
%    \begin{tabular}{l|c c c | c c c | c c c |c}
%         & LDA\cite{dirac_1930}& SVWN3\cite{PhysRevLett.101.266106}&SVWN5\cite{dirac_1930,SVWN5_1,1980CaJPh..58.1200V} & BLYP\cite{PhysRevA.38.3098,PhysRevB.37.785} &BP86\cite{PhysRevA.38.3098, PhysRevB.33.8822} & PBE\cite{PBE}& B3LYP\cite{PhysRevB.313.8822,  1993JChPh..98.1372B, 1993JChPh..98.5648B}   &PBE0\cite{1996JChPh.105.9982P,1999JChPh.110.6158A} & PBE38\cite{10.1063/1.3382344}& Ref.DT\cite{Dailey_JCP1949} \\
%         \hline
%         \hline
%         \ce{Cl}  & 0& 0 & 0&0 & 0& 0&  0& 0&0 & -110.4 \\
%         \ce{ICl}&-92.8 & -93.0 &-92.9 & -91.6& -91.4 &-90.8 &-92.3  & -91.8& -92.2&  -82.5 \\
%         \ce{ClCN}&-83.8 & -84.1 &-83.8 &-85.4 & -83.8& -83.2&-87.6  & -85.9&-87.1 &   -83.2\\
%         \ce{CH3Cl}&-78.1 & -78.3 &-78.1 &-77.9 &-77.1  &-77.1 & -77.9&-77.0 &-77.0 &   -75.13\\
%         \ce{NaCl}& -4.7&  -4.8&-4.7 &-4.6 &-4.9 &-4.7 &-5.1 &-5.3 & -5.5&   $<$1\\
%    \end{tabular}
%    \caption{Calculation of the nuclear quadrupole coupling constant for \ce{^35Cl} in the chlorine compounds present in Ref.~\citenum{Dailey_JCP1949}.}
%    \label{tab:my_label}
%\end{table}

\appendix
\section{Sternheimer shielding}\label{app:Sternheim}
In a 1950 letter to the editor, Sternheimer writes\cite{Sternheimer_PhysRev.80.102.2}
\begin{quotation}
It was pointed out by Rabi that the hyperfine splitting due to the
nuclear quadrupole moment includes the effect of an electric quadrupole
moment induced in the electron shells.
\end{quotation}
He then proceeds, using the Thomas-Fermi model, to provide ``as crude
estimate of the moment induced in a core of closed shells''. In a
1951 follow-up paper Sternheimer extends his analysis to the Hartree-Fock
level.\cite{Sternheimer_PhysRev.84.244} It is clear that the calculations
reported in the present paper does not feature the variational inclusion
of a nuclear quadrupole moment, and so one may worry if some effect
is missing. This is analyzed in the following.

The general electrostatic interaction of two charge densities $A$
and $B$ is expressed as
\[
E_{AB}=\frac{1}{4\pi\varepsilon_{0}}\int d^{3}\boldsymbol{r}_{1}\int d^{3}\boldsymbol{r}_{2}\frac{\rho_{A}\left(\boldsymbol{r}_{1}\right)\rho_{B}\left(\boldsymbol{r}_{2}\right)}{\left|\boldsymbol{r}_{1}-\boldsymbol{r}_{2}\right|}
\]
Using a Laplace expansion, we obtain
\begin{equation}
E_{AB}=\frac{1}{4\pi\varepsilon_{0}}\sum_{\ell=0}^{\infty}\sum_{m=-\ell}^{+\ell}\frac{4\pi}{2\ell+1}\int d^{3}\boldsymbol{r}_{1}\int d^{3}\boldsymbol{r}_{2}\frac{r_{<}^{\ell}}{r_{>}^{\ell+1}}Y_{\ell m}\left(\boldsymbol{n}_{1}\right)Y_{\ell m}^{\ast}\left(\boldsymbol{n}_{2}\right)\rho_{A}\left(\boldsymbol{r}_{1}\right)\rho_{B}\left(\boldsymbol{r}_{2}\right).\label{eq:genint}
\end{equation}
When $r_{2}>r_{1}$, a spherical electric multipole expansion is obtained
\begin{equation}
E=\sum_{\ell=0}^{\infty}\sum_{m=-\ell}^{+\ell}q_{\ell m}^{\ast}{\cal E}_{\ell m};\quad\begin{array}{lcl}
q_{\ell m} & = & \sqrt{\frac{4\pi}{2\ell+1}}\int r_{2}^{\ell}Y_{\ell m}\left(\boldsymbol{n}_{1}\right)\rho_{A}\left(\boldsymbol{r}_{1}\right)d^{3}\boldsymbol{r}_{1}\\
{\cal E}_{\ell m} & = & \frac{1}{4\pi\varepsilon_{0}}\sqrt{\frac{4\pi}{2\ell+1}}\int\frac{1}{r_{1}^{\ell+1}}Y_{\ell m}\left(\boldsymbol{n}_{2}\right)\rho_{B}\left(\boldsymbol{r}_{1}\right)d^{3}\boldsymbol{r}_{2}
\end{array};\quad r_{2}>r_{1}.
\end{equation}
Focusing on the electric quadrupole component $\ell=2$ and $m=0$
we get
\begin{equation}
E_{20}=q_{20}^{\ast}{\cal E}_{20};\quad\begin{array}{lcl}
q_{20} & = & \int\frac{1}{2}\left(3z_{1}^{2}-r_{1}^{2}\right)\rho_{A}\left(\boldsymbol{r}_{1}\right)d^{3}\boldsymbol{r}_{1}\\
{\cal E}_{20} & = & \displaystyle{\frac{1}{4\pi\varepsilon_{0}}\int\frac{\left(3z_{2}^{2}-r_{2}^{2}\right)}{2r_{2}^{5}}\rho_{B}\left(\boldsymbol{r}_{2}\right)d^{3}\boldsymbol{r}}_{2}
\end{array};\quad r_{2}>r_{1},
\end{equation}
where we used that
\begin{equation}
Y_{20}\left(\theta,\phi\right)=\sqrt{\frac{5}{4\pi}}\frac{1}{2}\left(3\cos^{2}\theta-1\right).
\end{equation}

Let us now investigate the Sternheimer shielding effect. We consider
an atom $A$, placed at the origin, with a nuclear quadrupole moment
Q, associated with the operator
\begin{equation}
\hat{h}_{Q}=-\frac{e^{2}Q}{4\pi\varepsilon_{0}}\frac{\left(3\cos^{2}\theta-1\right)}{4r^{3}};\quad eQ=2q_{20}.\label{eq:h_Q op}
\end{equation}
 The induced quadrupole moment can be calculated as a linear response
function
\begin{equation}
e\delta Q=2\delta q_{20}=2\left\langle \left\langle -\frac{e}{2}\left(3z^{2}-r^{2}\right);\hat{h}_{Q}\right\rangle \right\rangle .\label{eq:Qind_lr}
\end{equation}
We have investigated this for the {[}\ce{Mg}{]} core of the aluminum
atom. We first did an average-of-configuration (AOC) HF calculation\cite{thyssen:phd} 
of the aluminium atom at the non-relativistic level using the Lévy-Leblond
Hamiltonian,\cite{saue:spinfree,levy:eq} a dyall.ae3z basis\cite{dyall:H--Ar,dyall:bases} and the DIRAC code.\cite{DIRAC24} We next calculated
the above linear response function Eq.~\eqref{eq:Qind_lr}, but not allowing
the \ce{Al} $3p$ orbitals to respond. We find that the total induced
quadrupole moment is $e\delta Q$=-0.820E-01$Q$, which increases
significantly to -1.027$Q$ at the uncoupled HF-level. The uncoupled
linear response function can be exactly decomposed into individual
orbital contributions, as seen in Table \ref{tab:Induced-Al}. At
the coupled HF-level, reponse can be restricted to selected occupied
orbitals, but the contributions no longer sum up to the total value.
In both cases, it is seen that the response from $s$ and $p$ orbitals
is shielding (+) and anti-shielding (-), respectively. Finally, it
is possible to separate the contribution from $2p$ into excitations
to virtual $p$ and $f$ orbitals, as also shown in Table \ref{tab:Induced-Al}.
\noindent \begin{center}
\begin{table}
\noindent \begin{centering}
\begin{tabular}{clcc}
 &  & coupled & uncoupled\tabularnewline\hline
$\delta Q$ &  & -0.082 & -1.027\tabularnewline\hline
 & $1s$ & 0.058 & 0.056\tabularnewline
 & $2s$ & 0.161 & 0.144\tabularnewline
 & $3s$ & 0.631 & 0.429\tabularnewline
 & $2p$ & -2.166 & -1.656\tabularnewline
 & $2p\rightarrow p$ & -1.288 & -1.083\tabularnewline
 & $2p\rightarrow f$ & -0.682 & -0.573\tabularnewline
\end{tabular}
\par\end{centering}
\caption{Induced quadrupole moment (in units of Q) of the aluminium atom, calculated
at the non-relativistic HF level using the Lévy-Leblond Hamiltonian,
a dyall.ae3z basis and the DIRAC code.\label{tab:Induced-Al}}

\end{table}
\par\end{center}

We have seen that the quadrupole moment induced in the electronic
cloud may be of the same order as the nuclear one. We now investigate
its effect on energies. In his first two papers, Sternheimer investigated
open-shell atoms. It is clear that the expectation value of the spherically
symmetric closed core shells with respect to the operator, Eq.~\eqref{eq:h_Q op}, 
is zero; any non-zero contribution comes from the open-shell valence,
which now may also interact with the polarized core. The operator
form Eq.~\eqref{eq:h_Q op} assumes a point-like nuclear quadrupole moment.
The induced quadrupole moment
\[
e\delta Q=2q_{20}=-e\int\left(3z_{2}^{2}-r_{2}^{2}\right)\delta_{Q}n\left(\boldsymbol{r}_{2}\right)d^{3}\boldsymbol{r}_{2}
\]
is certainly not point-like and so we should use the general interaction
expression, Eq.~\eqref{eq:genint}, picking out the contribution from $\ell=2$
and $m=0$. It may be written as
\begin{equation}
E_{A\delta Q}=-\frac{e}{4\pi\varepsilon_{0}}\int d^{3}\boldsymbol{r}_{1}\rho_{A}\left(\boldsymbol{r}_{1}\right)\frac{\left(3z_{1}^{2}-r_{1}^{2}\right)}{4r_{1}^{5}}\gamma\left(r_{1}\right)Q\label{eq:E_A_dQ}
\end{equation}
where appears the Sternheimer \emph{screening} factor \cite{Sternheimer_Foley_PhysRev.92.1460}
\[
\gamma\left(r_{1}\right)=\left(1/Q\right)\left[\int_{r_{2}<r_{1}}d^{3}\boldsymbol{r}_{2}\left(3z_{2}^{2}-r_{2}^{2}\right)\delta_{Q}n\left(\boldsymbol{r}_{2}\right)+r_{1}^{5}\int_{r_{2}>r_{1}}d^{3}\boldsymbol{r}_{2}\frac{\left(3z_{2}^{2}-r_{2}^{2}\right)}{r_{2}^{5}}\delta_{Q}n\left(\boldsymbol{r}_{2}\right)\right].
\]
We note that the total induced quadrupole moment can be expressed
as
\[
\delta Q=\gamma_{\infty}Q;\quad\gamma_{\infty}=\lim_{r\rightarrow\infty}\gamma\left(r\right).
\]
In Eq.~\eqref{eq:E_A_dQ} $\rho_{A}$ refers to whatever non-symmetric
charge distribution is in interaction with the induced quadrupole
moment, for instance the open-shell valence in an atom. 

In subsequent papers\cite{Sternheimer_Foley_PhysRev.92.1460,Foley_PhysRev.93.734},
Sternheimer and co-workers moved on to the consideration of diatomic
molecules. In the first paper, devoted to the \ce{Li2} molecule, we
find the statement
\begin{quote}
In order to obtain the shielding effect without ambiguity, it is best
to study this effect as a result of the distortion of the $1s$ shell
by the asymmetric potential caused by the other charges. The correction
to \emph{q} is the due to the part of the distortion which behaves
as $\left(3\cos^{2}\theta-1\right)$. This picture is equivalent to
the consideration of the induced moment, as will now be shown.
\end{quote}
The demonstration is somewhat convoluted and we therefore give a simplified
version here. We follow the second paper, considering the highly polar
\ce{NaCl} molecule. We investigate the effect of a nuclear quadrupole
moment of the chlorine nucleus at the origin and represent the \ce{Na+} moiety 
as a point charge $+e$ at the position $\boldsymbol{R}_{\text{Na}}=\left(0,0,R\right)$
well outside the electron cloud of the \ce{Cl-} moiety. The interaction
energy Eq.~\eqref{eq:E_A_dQ} can then be written as
\begin{equation}
E_{\text{Na},\delta Q_{Cl}}=-\frac{e^{2}}{4\pi\varepsilon_{0}}\frac{\gamma_{\infty}Q}{2R^{3}}=-\frac{e^{2}}{4\pi\varepsilon_{0}}\frac{1}{2R^{3}}\int d^{3}\boldsymbol{r}\left(3z^{2}-r^{2}\right)\delta_{Q}n\left(\boldsymbol{r}\right).\label{eq:E_NaCl}
\end{equation}

The induced charge distribution can itself be expressed as a linear
response function 
\[
\delta n_{Q}\left(\boldsymbol{r}\right)=\left\langle \left\langle \delta\left(\boldsymbol{r}\right);\hat{h}_{Q}\right\rangle \right\rangle .
\]
For the present purposes, it suffices to look at the linear response
function at the uncoupled HF-level. Starting from a perturbation operator
on the form 
\[
V=\sum_{X}\varepsilon_{X}\hat{h}_{X},
\]
the linear response function is then given by
\begin{equation}
\left\langle \left\langle \hat{h}_{A};\hat{h}_{B}\right\rangle \right\rangle =\left[\frac{d^{2}E}{d\varepsilon_{A}d\varepsilon_{B}}\right]_{\boldsymbol{\varepsilon}=\boldsymbol{0}}=\sum_{ai}\left[\frac{\langle\Phi_{0}|\hat{h}_{A}|\Phi_{i}^{a}\rangle\langle\Phi_{i}^{a}|\hat{h}_{B}|\Phi_{0}\rangle}{\varepsilon_{i}-\varepsilon_{a}}+\text{c.c.}\right]=\left\langle \left\langle \hat{h}_{B};\hat{h}_{A}\right\rangle \right\rangle .\label{eq:linear_response}
\end{equation}
In the present case we obtain
\[
\delta n_{Q}\left(\boldsymbol{r}\right)=\sum_{ai}\frac{1}{\varepsilon_{i}-\varepsilon_{a}}\left[\varphi_{i}^{\ast}\left(\boldsymbol{r}\right)\varphi_{a}\left(\boldsymbol{r}\right)\langle\Phi_{i}^{a}|\hat{h}_{Q}|\Phi_{0}\rangle+\text{c.c.}\right]
\]
 The interaction energy therefore reads
\begin{equation}
E_{\text{Na},\delta Q_{Cl}}=-\frac{e^{2}}{4\pi\varepsilon_{0}}\frac{1}{2R^{3}}\sum_{ai}\frac{1}{\varepsilon_{i}-\varepsilon_{a}}\int d^{3}\boldsymbol{r}_{2}\left(3z_{2}^{2}-r_{2}^{2}\right)\left[\varphi_{i}^{\ast}\left(\boldsymbol{r}_{2}\right)\varphi_{a}\left(\boldsymbol{r}_{2}\right)\langle\Phi_{i}^{a}|\hat{h}_{Q}|\Phi_{0}\rangle+\text{c.c.}\right].\label{eq:E_NaCl_2}
\end{equation}

Let us now consider the distortion of the \ce{Cl-} moiety by the
sodium charge. The associated interaction operator is 
\[
\hat{h}_{\text{Na}}=-e\phi_{\text{Na}}=\frac{-e^{2}}{4\pi\varepsilon_{0}}\frac{1}{\left|\boldsymbol{r}-\boldsymbol{R}_{\text{Na}}\right|}
\]
 However, we shall be interested only in the component associated
with $\ell=2$ and $m=0$, which is
\[
\hat{h}_{\text{Na},E2}=\frac{-e^{2}}{4\pi\varepsilon_{0}}\frac{1}{2}\left(3z^{2}-r^{2}\right)R^{-3}
\]
 Returning to the above interaction energy, Eq.~\eqref{eq:E_NaCl_2}, we
find that it can be expressed as a linear response function
\[
E_{\text{Na},\delta Q_{Cl}}=\left\langle \left\langle \hat{h}_{\text{Na},E2};\hat{h}_{Q}\right\rangle \right\rangle .
\]
Such a response function gives the modification of expectation value
of $\hat{h}_{P,E2}$ under the influence of $\hat{h}_{Q}$, but equally
well the opposite, as clearly seen from Eq.~\eqref{eq:linear_response}.
The statement by Sternheimer and co-workers has therefore been shown
to hold true. Our SCF calculations do not include the operator Eq.~\eqref{eq:h_Q op}
associated with a nuclear quadrupole moment, but do generate a fully
relaxed electron cloud, which enters the perturbative calculation
of their interaction. The Sternheimer shielding is therefore fully
accounted for.

\section{Localization and core orbitals}\label{app:loc}
Orbital localization in the DIRAC code is based on the Pipek--Mezey criterion.\cite{Pipek-Mezey_JCP1989} A first implementation was based on the steepest ascent method.\cite{Dubillard_JCP2006} Building upon the work of Jørgensen and co-workers,\cite{Jansik_JCP2011,Hoyvik_JCTC2012} we have next implemented localization of orbitals using the trust-region algorithm of Fletcher,\cite{Fletcher_opt} in particular as outlined in Ref. \citenum{helgaker:bok} (Section 12.3). Our algorithm takes into account Kramers pairing due to time-reversal symmetry. 

The Pipek–Mezey localization criterion minimizes the number of centers spanned by the individual MOs. However, in practice, we see that our localization procedure touches core orbitals, which are maximally localized. 
In the following we first specify the exponential parametrization of our algortim, allowing unconstrained optimization as well as the straightforward identification and elimination of redundancies. We then derive expression for the gradient and Hessian, and finally address the question as to whether the Pipek--Mezey algorithm may modify core orbitals.

\subsection{Basic theory}
We use the localization criterion of Pipek and Mezey,\cite{Pipek-Mezey_JCP1989} where the localized orbitals are obtained by maximization of the functional (inverse mean delocalization)
\begin{equation}
    G\left[\left\{ \varphi_{i}\right\} \right]=\frac{1}{N}\sum_{i=1}^{N}\sum_{A}\left(Q_{i}^{A}\right)^{2};\quad Q_{i}^{A}=\langle\varphi_{i}|P_{A}|\varphi_{i}\rangle.
\end{equation}
 The quantity $Q_{i}^{A}$ is the contribution from MO with index
$i$ to the Mulliken charge of atom $A$ and is a diagonal element 
of the Hermitian matrix
\begin{equation}\label{Q_ij}
    Q_{ij}^{A}=\boldsymbol{c}_{i}^{A\dagger}S^{AA}\boldsymbol{c}_{j}^{A}+\frac{1}{2}\sum_{B\ne A}\left(\boldsymbol{c}_{i}^{A\dagger}S^{AB}\boldsymbol{c}_{j}^{B}+\boldsymbol{c}_{i}^{B\dagger}S^{BA}\boldsymbol{c}_{j}^{A}\right),
\end{equation}
The expression assumes that the overlap matrix $S$ in the atomic basis is sorted on atomic centers, so that $S^{AB}$ is an off-diagonal block associated with centers $A$ and $B$. The vector $\boldsymbol{c}_{i}^{A}$ collects expansion coefficients from center $A$ for orbital of index $i$. Obviously,
\begin{equation}
    \sum_{A}Q_{ij}^{A}=\delta_{ij}.\label{eq:Qsum}
\end{equation}

To set up our second-order algorithm, we introduce an exponential
parametrization
\begin{equation}
\tilde{\varphi}_{p}=\sum_{q}\varphi_{q}U_{qp};\quad U=\exp\left[-\kappa\right];\quad\kappa_{pq}^{\ast}=-\kappa_{qp}.\label{eq:parametrization}
\end{equation}
Let us first investigate redundancies. In the ensuing manipulations, it will be helpful to expand the Pipek--Mezey functional in orders of the orbital rotation parameters $\{\kappa_{pq}\}$. For e.g. the Hartree--Fock method in second quantization, this is achieved using the Baker--Campbell--Haussdorf expansion.\cite{helgaker:bok} We obtain a suitable expansion in the present case starting from
\begin{equation}
G\left(\lambda\right)=\sum_{i}\sum_{A}\sum_{pqrs}\langle\varphi_{p}\left\{ \exp\left[-\lambda\kappa\right]\right\} _{pi}|P_{A}|\varphi_{q}\left\{ \exp\left[-\lambda\kappa\right]\right\} _{qi}\rangle\langle\varphi_{r}\left\{ \exp\left[-\lambda\kappa\right]\right\} _{ri}|P_{A}|\varphi_{s}\left\{ \exp\left[-\lambda\kappa\right]\right\} _{si}\rangle.
\end{equation}
Next we consider the first derivative of $G\left(\lambda\right)$
at $\lambda=0$
\begin{equation}
    G^{\prime}\left(0\right)= -2\sum_{A}\sum_{i}\sum_{j}\left(\kappa_{ji}^{\ast}Q_{ji}^A+\kappa_{ji}Q_{ij}^A\right)Q_{i}^{A}
\end{equation}
We first note that for the contribution $i=j$, we get
\begin{equation}
    Q_{i}^{A}\left(\kappa_{ii}^{\ast}+\kappa_{ii}\right)=0.
\end{equation}
since non-zero diagonal elements of an anti-Hermitian matrix can only
be purely imaginary. We therefore conclude that diagonal elements
of the $\kappa$ matrix are redundant. From the matrix symmetry, it
is clear that we only need one triangle, and we choose the lower one,
that is $\left\{ \kappa_{ij,}\kappa_{ij}^{\ast},i>j\right\}$. 

The parameter space is navigated starting from a quadratic expansion of the Pipek--Mezey function
\begin{equation}\label{eq:Gmod}
    \tilde{G}(\{\mathbf{\kappa}\})=G{[0]}+G^{[1]\dagger}\mathbf{\kappa}+\frac{1}{2}\mathbf{\kappa}^{\dagger}G^{[2]}\mathbf{\kappa}
\end{equation}
Elements of the gradient $G^{[1]}$  are:
\begin{equation}
\displaystyle{\left[\frac{\partial G}{\partial\kappa_{ij}^{\ast}}\right]_{\boldsymbol{\kappa}=\boldsymbol{0}}} = 2\displaystyle{\sum_{A}}Q_{ij}^{A}\left(Q_{i}^{A}-Q_{j}^{A}\right);\quad i>j.\label{eq:gradient}
\end{equation}
Elements of the Hessian $G^{[2]}$ are ($i>j$, $k>l$):
\begin{eqnarray}
\left[\frac{\partial^{2}G}{\partial\kappa_{ij}^{\ast}\partial\kappa_{kl}}\right]_{\boldsymbol{\kappa}=\boldsymbol{0}} & = & 2\sum_{A}Q_{ij}^{A}\left(\delta_{ik}Q_{li}^{A}-\delta_{il}Q_{ik}^{A}+\delta_{jl}Q_{jk}^{A}-\delta_{jk}Q_{lj}^{A}\right)\\
 & + & \sum_{A}\left(\delta_{ik}Q_{lj}^{A}\left(2Q_{i}^{A}-Q_{j}^{A}-Q_{l}^{A}\right)+\delta_{jl}Q_{ik}^{A}\left(2Q_{j}^{A}-Q_{i}^{A}-Q_{k}^{A}\right)\right)\\
 \left[\frac{\partial^{2}G}{\partial\kappa_{ij}^{\ast}\delta\kappa_{kl}^{\ast}}\right]_{\boldsymbol{\kappa}=\boldsymbol{0}}& = & 2\sum_{A}Q_{ij}^{A}\left(\delta_{ik}Q_{il}^{A}-\delta_{il}Q_{ki}^{A}+\delta_{jl}Q_{kj}^{A}-\delta_{jk}Q_{jl}^{A}\right)\\
 & - & \sum_{A}\left(\delta_{jk}Q_{il}^{A}\left(2Q_{j}^{A}-Q_{l}^{A}-Q_{i}^{A}\right)+\delta_{il}Q_{kj}^{A}\left(2Q_{i}^{A}-Q_{j}^{A}-Q_{k}^{A}\right)\right).
\end{eqnarray}
The remaining combinations are obtained by complex conjugation. 
Within the validity of the quadratic model Eq.~\eqref{eq:Gmod}, the stationary point is found by taking the Newton step
\begin{equation}\label{eq:Newton}
    G^{[2]}\mathbf{\kappa}=-G^{[1]};
\end{equation}
beyond a level-shifted Newton step is determined to find the optimal step on the boundary of the trust region.\cite{helgaker:bok} 

\subsection{Will the Pipek--Mezey algorithm modify core orbitals ?}
For simplicity let us limit attention to single orbital rotation
parameter $\kappa_{ij}$ and two centers $A$ and $B$. Matrix elements, Eq.~\eqref{Q_ij}, reduce to 
\begin{equation}
    Q_{ij}^{A}=\frac{1}{2}\left(\boldsymbol{c}_{i}^{A\dagger}S^{AB}\boldsymbol{c}_{j}^{B}+2\boldsymbol{c}_{i}^{A\dagger}S^{AA}\boldsymbol{c}_{j}^{A}+\boldsymbol{c}_{i}^{B\dagger}S^{BA}\boldsymbol{c}_{j}^{A}\right)
\end{equation}
and the sum rule, Eq.~\eqref{eq:Qsum}, is now
\begin{equation}
    Q_{ij}^{A}+Q_{ij}^{B}=0;\quad Q_{i}^{A}+Q_{i}^{B}=1.
\end{equation}
The Newton step, Eq.~\eqref{eq:Newton}, can thereby be expressed as
\begin{equation}
    \left[\begin{array}{cc}
8\left|Q_{ij}^{A}\right|^{2}-4\left(Q_{i}^{A}-Q_{j}^{A}\right)^{2} & 8Q_{ij}^{A}Q_{ij}^{A}\\
8Q_{ij}^{A\ast}Q_{ij}^{A\ast} & 8\left|Q_{ij}^{A}\right|^{2}-4\left(Q_{i}^{A}-Q_{j}^{A}\right)^{2}
\end{array}\right]\left[\begin{array}{c}
\kappa_{ij}\\
\kappa_{ij}^{\ast}
\end{array}\right]=-\left[\begin{array}{c}
4Q_{ij}^{A}\left(Q_{i}^{A}-Q_{j}^{A}\right)\\
4Q_{ij}^{A\ast}\left(Q_{i}^{A}-Q_{j}^{A}\right)
\end{array}\right];\quad i\ne j
\end{equation}

 We clearly see that the gradient is zero if $Q_{i}^{A}=Q_{j}^{A}$
or $Q_{ij}^{A}=0$. 
Let us consider what happens when $\boldsymbol{c}_{i}^{B}=\boldsymbol{0}$
such that $Q_{i}^{A}=1$ for some specific nucleus $A$. The solution to the above Newton equation is then
\begin{equation}
    \left[\begin{array}{c}
\kappa_{ij}\\
\kappa_{ij}^{\ast}
\end{array}\right]=-\frac{1}{\mbox{det}G^{[2]}}\left[\begin{array}{cc}
8\left|Q_{ij}^{A}\right|^{2}-4\left(1-Q_{j}^{A}\right)^{2} & -8Q_{ij}^{A\ast}Q_{ij}^{A\ast}\\
-8Q_{ij}^{A}Q_{ij}^{A} & 8\left|Q_{ij}^{A}\right|^{2}-4\left(1-Q_{j}^{A}\right)^{2}
\end{array}\right]\left[\begin{array}{c}
4Q_{ij}^{A}\left(1-Q_{j}^{A}\right)\\
4Q_{ij}^{A}\left(1-Q_{j}^{A}\right)
\end{array}\right],
\end{equation}
where
\begin{equation}
    \mbox{det}G^{[2]}=16\left(1-Q_{j}^{A}\right)^{2}\left(\left(1-Q_{j}^{A}\right)^{2}-2\left|Q_{ij}^{A}\right|^{2}\right)
\end{equation}
These manipulations are not valid if the gradient is zero, but the result clearly suggests that the Pipek--Mezey algorithm may touch fully localized orbitals.
%\nocite{*}
\bibliography{article}% Produces the bibliography via BibTeX.

\end{document}